\documentclass[12]{article}
\usepackage{mathtext}
\usepackage[cp1251]{inputenc}
\usepackage[T2A]{fontenc}
\newcommand{\nnn}{\bigskip}

\renewcommand{\Im}{{\rm Im}}
\renewcommand{\d}{\delta}
\renewcommand{\b}{\beta}
\newcommand{\g}{\gamma}

\newcommand{\sign}{{\rm sign}}

\newcommand{\D}{\Delta}

\newcommand{\e}{\epsilon}

\newcommand{\ar}{\longrightarrow}
\newcommand{\w}{\omega}

\newcommand{\la}{\lambda}
\renewcommand{\a}{\alpha}
\begin{document}
\title{Algorithmic approach to quantum physics}
\author{Y.I.Ozhigov\thanks{e-mail: 
ozhigov@cs.msu.su} \\[7mm]
Moscow State University,\\
Institute of Physics and technology RAS
} 
\maketitle
\begin{abstract}
Algorithmic approach is based on the assumption that any quantum evolution of many particle system can be simulated on a classical computer with the polynomial time and memory cost. Algorithms play the central role here but not the analysis, and the simulation gives the "film" which visualizes many particle quantum dynamics and is demonstrated to a user of the model. Restrictions following from the algorithm theory are considered on a level of fundamental physical laws. Born rule for the calculation of quantum probability as well as the decoherence is derived from the existence of a nonzero minimal value of amplitude module - a grain of amplitude. The limitation on the classical computational resources gives the unified description of quantum dynamics that is not divided to the unitary dynamics and measurements and does not depend on the existence of observer. It is proposed the description of states based on the nesting of particles in each other that permits to account the effects of all levels in the same model. Algorithmic approach admits the possibility of refutation, because it forbids the creation of a scalable quantum computer that is allowed in the conventional quantum formalism. 

\end{abstract}

\section{Introduction}

The notion of classical algorithm and computational methods headily penetrate to all areas of natural sciences. This penetration gives the new language for the description of science that is based rather on algorithms than on formulas and principles as the conventional approaches. This new approach brings the serious change of the contents of sciences that is not yet fully realized due to the incredible flexibility and universality of the algorithmic description of Nature. 
But we already meet with the surprising features of this new description that distinguish it from the conventional language and these features can be in principal established in experiments. It is connected with the most advanced part of the natural science - physics, or more precisely, quantum physics where this difference has been revealed very explicitly; it is just the subject of this article\footnote{Several researchers have come to the idea of this approach independently; here I mention V.Akulin who expressed it in the talk with the author; some suggestions of the limitation on the area of applicability of quantum formalism are permanently expressed by the other scientists, especially by those who deal with the different aspects of the decoherence problem (see, for example, (\cite{Fe},\cite{Ak}).}. 

The algorithmic approach to physics is based on the simple idea: a computer must be considered as a paramount physical device which necessarily attends at any experiments. It was just so even when there were no computers, their role was played by a physicist who fulfilled all the computations at hand. It follows from this assumption that all the limitations that result from algorithm theory must be considered equally to the physical laws. We call these limitations computational. The main assumption of algorithmic approach was not very pithy when the computational limitations could be ignored, e.g., when the existence of a computer could be neglected or it could be used as a simple calculating machine for calculations by fixed formulas\footnote{Strictly speaking it is never possible to ignored the computational limitations. For example, even exact mathematical consideration of one electron movement in vacuum with the account of relativistic effects (transformations of photons and electron-positron pairs) leads to the summing of divergent sums and several tricks, for which the mathematical substantiation is based on the implicit confession of the priority of algorithms over the descriptive technique like particles and their interactions. Just such a priority is used in the method of renormalization where the classical viewpoint to the full consideration of space-time is sacrificed to the preserving of the convenience of the algorithmic description of dynamics. Quantum method won in atomic physics just because it gave the effective algorithms leading to the right predictions as opposed to classical method. For example, Shroedinger equation gives the hydrogen atom spectrum after easy computations whereas the classical method without neglecting electromagnetism leads to the wrong prediction, and the classical consideration of the electron movement in electromagnetic field based on Maxwell equations and relativistic formula for energy has not yet fulfilled because this problem statement gives no effective algorithms.}.
The situation has been changed when the traditional analytical formalism of physics came into conflict with its new problems. It became evident with the beginning of the elaboration of a hypothetic quantum computer, which was appealed to in order to resolve this conflict. The concept of a quantum computer proceeds from the assumption of the priority of analytical technique over algorithms. E.g., here it is implicitly assumed that the concept of a classical algorithm is not fundamental and can be easily replaced by the computational procedure of the other type - quantum computation. This assumption is the attempt to extrapolate quantum physics to the area where it has been never checked, hence a quantum computer must be treated as a principal hypothesis to which we will return later. 

Now we try to understand what we can obtain with algorithmic approach, when we assume the concept of algorithm as the most fundamental. 
The key assumption of this approach is the theoretical possibility to create the complete model of observed events by means of a classical computer which uses effective algorithms and is independent of an experimentalist. An effective algorithm is such a classical algorithm that requires the quantity of resources (time and memory) limited by some polynomial of the memory size needed for the description of the simulated system states\footnote{This class of algorithms is independent from the formalization of algorithms.}. 
Such a model must show in each moment the distribution of the probability to find any subsystem of the simulated system in any possible state for which this probability $p$ is sufficiently large. For example we can assume that
$pT_{tot}>1$, where $T_{tot}$ - is the largest accessible value of the time. 

The work of such an algorithm can be represented in the form of a "film" which is demonstrated to a user, who cannot interfere to this "film"; a user can only order such a "film" beforehand and point, for example, what measurements and when it is planned to do over the simulated system\footnote{Of course, if we allow to a user to interfere in such a "film" the problem would become insoluble due to the quantum non-locality. But just our problem statement has the practical significance because it can answer to the main question of an experimentalist: what these or that actions over the given system result in. Here the delay that compensates the lack of processing speed cannot in turn exceed the established limits. Practically, the time of simulation must grow not faster than linearly as the size of simulated system grows, because otherwise we cannot hope to create even the film with minimum substance.}.
In other words we consider the Nature as the "film" demonstrated to us through the simulating computer that is included to the computational network, and as users we have no rights to access inside this network. Such a network can be considered as a model of the frame accosiated with the given user.
All the information this computer deals with has thus the form of binary strings of the limited length. 
The physical magnitudes needed for the right demonstration of the "film" (for example, binding energy in a molecule, the mean distance between atoms etc.) are computed by the simulating computer in course of the "film" preparation and are used in its demonstration. Such "films" are the most general form of the physical phenomena description corresponding to the algorithmic approach. The matter thus concerns the replacement of the conventional mathematical apparatus of physics (analysis and algebra) to the different mathematical apparatus (algorithms) that is more general but by virtue of the historical reasons is less known to physicists. We will not develop this topic here and concentrate on the practical side of this approach. We only note that the advantage of this description is that the specialists in different areas can work with it independently; this possibility itself can be crucial for the future of this approach.

The principal consequence of algorithmic approach is the existence of the minimal and nonzero value of amplitudes molude, the so called amplitude quantum (or grain). The thesis about amplitude quantum gives such a classical urn scheme for quantum probability that implyies Born rule (see below). Moreover, the concept of amplitude quantum makes possible to give a unified description of quantum dynamics which is not divided to the unitary dynamics and measurements and does not depend on the existence of observer. It makes the "film" representation of dynamics as objective as the conventional representation by formulas. 

Algorithmic approach which we are going to consider arises from the attempts to create a 
computer model for the dynamics of many particle systems with quantum behavior, for example, chemical reactions. It includes the dynamics of atoms and molecular structures connected with the  change of electronic states inducing the creation of chemical bounds between atoms. 
It is well known that the behavior of an electron cannot be described in terms of classical dynamics, say by the representation of it as a ball moving in the Coulomb potential of the nucleus. All the more it is impossible for a system of several electrons. 
The principal difficulty arises already for the states of many electrons in atoms and molecules. The dimensionality of the space of these states grows exponentially when the number of electron increases. 
The exponential growth takes place even if we limit the number of exited one electron levels by a linear function. 
Such states are usually represented by Fock-Sleter determinants composed from one electron functions which are chosen from the condition of zero energy variation of many electron system (see (\cite{Sl})). When computing such determinants we have to fulfill exponential work depending on its size and their total number will grow exponentially as well. This is why existing algorithms of molecular simulation account the number of electrons  limited beforehand (for example, two electrons for each valence bound only, and even for these two electrons the computation of state is fulfilled  not in the whole space but in the approximation of mean field or the similar). 
Quantum states of nuclei are not taken into account at all, the nuclei are considered as "balls with the springs" where the "springs" are determined by the stationary electronic configurations and the Coulomb interactions between nuclei. 
The corrections connected with the quantum character of the nuclei movement can be then introduced to such a model by hand. For example, the well known and very important phenomenon of a proton tunneling  requires the quantum description in the form of wave function, not by the classical way; the tunneling of a nitrogen atom in a molecule of ammonia that results in the observed spectrum of this molecule, hydrogen bounds etc.  
 This type of more complex phenomena cannot be described in terms of "balls with springs", but this model yet can be in principle modified to account independent tunneling of separated nuclei. But there exist more complex phenomena connected with the quantum entanglement between electrons and nuclei and between nuclei. 
The diffraction of a molecule on a slit represents the simplest example of such entanglement. Here the whole molecule behaves as a single quantum particle. Such a phenomenon in principal cannot be simulated by the method of mean field or "balls and springs". These phenomena are called "collective excitations". The known attempts to simulate such movements are based on serious limitations of the movements of particles in such systems. For example, in the work (\cite{NF}) it is assumed that the particles are represented by the separated Gaussian wave packages and thus this way does not give the universal method of the simulation of many body quantum systems. 

One more type of effects that are beyond the area of classical simulation methods is connected with the electrodynamics. The effect of delay of an electromagnetic field action on a slow charged particle and the other relativistic effects can be always neglected in the computation of atomic spectra. For example, the Lamb shift of energy levels is of the order $10^{-3}$ from the difference between the nearest levels. But in the chemical reactions it is not admissible to neglect the effect of emitting and absorption of photons, for example, in photosynthesis it plays the key role; there exist the methods of control over chemical reactions by laser impulses (see (\cite{SB},\cite {FTK}). 

The necessity to account the quantum nature of elementary particles in the study of molecular transformations was realized long ago, but it is difficult to do it practically due to the principal differences between quantum and classical forms of dynamics description. The main difficulty in classical many body dynamics arises from the instability of trajectories. It results in the chaotic behavior of the system\footnote{The conventional way here is connected with the application of various tricks based on the probability theory, for example, thermodynamics.}. This difficulty disappears if we assume that the set of values of coordinates and speeds (in the classical case it is the space of all one particle states) is grained as always is assumed in the computer simulation. In quantum case this assumption cannot resolve the problem because the main difficulty is another. Here the space of one particle states is not the set of all values for coordinates (or speeds) but is a linear combination of all such values. It is impossible to reduce such a state to one value of coordinate and speed because of the uncertainty principle: the more exact value of the coordinate we know the bigger dispersion in speed we shall have and vice versa. At the same time for the dynamical description we must know the coordinate and speed of a particle (in quantum case - the amplitude distribution among all coordinates or speeds). The necessity to use linear combinations of all values of the dynamical variables is the principal and remains also if we assume that the set of such values is grained. It leads to the exponential growth of the dimensionality of many particle state spaces that is the main obstacle in the quantum case. 
Just because of this obstacle the probability methods cannot be the main tool in the simulation of many body quantum systems. 

Simulation of many body quantum systems has several features that differs it from the other problems of theoretical physics and that reveals the weakness of the conventional analytical formalism of quantum mechanics. Here the integral picture is necessary that includes not only unitary segments of evolution but also sequential measurements which must be treated as independent from the existence of observer that is hardly compatible with the conventional analytical quantum formalism. The simulation based on the analytical formalism thus requires ceaseless switches from the quantum description to the classical and vise versa. 
The second difficulty is the exponential growth of the space of states dimensionality in the quantum simulation. This difficulty makes the problem of algorithmic description for the many body quantum systems the fundamental scientific problem, because it raises the question: how our world is designed, does it allow the effective classical algorithmic description or not. Of course, this question in its philosophical form is known for a long time - at least since the formalization of algorithm. But after the invention of a Quantum Computer (QC) this question has turned into the concrete scientific problem which presumes the certain solution. The point is that there is the clear procedure of verification: is a given device a QC or not. If the Hilbert spaces of exponential dimensionality are an adequate formalism then we have a principal possibility to create a scalable QC. This, still hypothetic device could solve some computational tasks substantially faster than any possible classical computer. For example, the problem of the factorization of integers can be solved by QC with almost exponential speedup (see (\cite{Sh})), the search problem - with quadratic speedup (see (\cite{Gr})). It is very important that QC is able to solve many particle Shroedinger equation in the time of order $t^2$, where $t$ is the physical time, or in other words it can simulate quantum many body dynamics without any simplifications!\footnote{The idea of QC was put forward by Feynman, and also Benioff and some others in order to give the new principal way for the simulation of many body systems. This guess became the exact result in the works (\cite{Za}) and (\cite{Wi}).}
The building of a scalable QC would mean the bankruptcy of algorithmic approach because no effective classical algorithm can simulate the work of QC. Really, if such a simulation is possible we would obtain the classical algorithm that solves all search problems substantially faster than by brute force that is impossible\footnote{Strictly speaking it is not the established mathematical theorem but only generalization of that can be called "mathematical practice", e.g. a meta-mathematical proposition which can obtain the exact form if we oversimplify it (the Church-Turing principle represents a remote association). But the conclusions of such "practice" are usually assumed in physics without objections. The reason is that such oversimplification does not stretch beyond the frameworks of usual abstraction which is used in the transfer from the natural phenomena to the mathematical formalism. The transfer to algorithmic approach just is the replacement of one type of formalism (analysis) to the other (algorithms).}.
I will not discuss here the condition of experimental works in quantum computing that yet have not shown the evident success (one can address to the general electronic archive http://xxx.lanl.gov). 
The development of QC technologies is absolutely necessary direction in quantum physics that logically follows from the conventional formalism. The logical chain: "algebraic technique - many body wave function - QC" is absolutely fundamental. 
The more detailed investigations of the existing approaches to the creation of QC as solid state quantum dots, ion traps and Josephson junctions (see (\cite{VK})) confirm the natural conclusion: there is no prohibition to the creation of a scalable QC in the known physics. But the progress in the experimental area goes too slowly that we can consider the prospects of the creation of QC as a single possibility. Just the absence of the clear advance in experiments is the main cause of the interest to the alternative - algorithmic approach that we are going to consider now.

Two moments arouse suspicion that the prohibition of the existing of a scalable QC can exist in Nature despite of that it cannot be derived from quantum mechanics. The first moment is connected with the decoherence that is treated as the influence of environment to a quantum system and leads to an irreversible corruption of its state. Just the decoherence is usually made responsible for the obstacles in the QC building. The short explanation of the sense of decoherence in the standard quantum mechanics looks as follows. 
Let the first qubit $|\psi\rangle_{sys}$ denote the state of considered system\footnote{Qubit is taken for the simplicity. It can be replaced by a wave vector in the space of states of arbitrary dimensionality.}, $|\phi\rangle_{env}$ - be the state of its nearest vicinity. If we initially prepared the system in a state $|\psi\rangle_{sys}=\a |0\rangle_{sys}+\b |1\rangle_{sys}$, and its vicinity was in an indifferent state$|0\rangle_{env}$, then it is reasonable to suppose that after the contact the extended system: "system + nearest vicinity" will be in the state which can be obtained from the initial state before contact $|\Psi_{ini}\rangle=(\a |0\rangle_{sys}+\b |1\rangle_{sys})\bigotimes |0\rangle_{env}$ by the application of some entangling operator, for example, CNOT, and it results in the entangled state $|\Psi_{fin}\rangle=(\a |0\rangle_{sys}\bigotimes |0\rangle_{env}+\b |1\rangle_{sys}\bigotimes |1\rangle_{env}$. If now the nearest vicinity will interact with its nearest vicinity, that in turn with its vicinity etc., we finally obtain the state of the form $\a |00\ldots 0\rangle+\b |11\ldots 1\rangle$ (we omit the sign of tensor product) where the dimensionality grows due to the permanent growth of the chain of vicinities denoted by dots. 
In this moment in the framework of standard quantum formalism it is assumed that the observation of one of these vicinities results in the collapse of the whole state and the initial system will be in one of the states $|0\rangle_{sys}$, or $|1\rangle_{sys}$, which means the action of decoherence. The weakness of this description is clear - it requires the presence of an observer that is not permissible in the simulation because it means the permanent artifact because an observer cannot be described in the framework of this formalism even if we have a QC. 
This is the irremovable feature of standard Hilbert formalism for systems of many particles and it is always simply ignored and replaced by the reference to the classical character of a measuring device (\cite{Fe}).

The second moment is that the exponential dimensionality of Hilbert spaces for quantum systems states has never been checked experimentally. All the facts approved in experiments and theoretically explained till nowadays can be derived from the theory by effective classical algorithms. It means that the real physics now lies within the bounds of effective algorithms, and there is no contradictory evidence to the universality of the algorithmic approach to physics. Practical methods of computations that reduce theoretical schemes to effective algorithms can be complicated but its heuristics is usually simple. For example in the computation of wave functions of stationary states of electrons in atoms the method of mean field is usually applied instead of the solution of many particle Shroedinger equation, e.g., we consider the behavior of one electron in the field induced by the others accordingly to the probability distribution of their coordinates. This method gives the satisfactory agreement with experiments, say it allows to find a good approximation of the energy of ionization, spectrum and spatial configurations of molecules. All more exact computations including, for example, relativistic corrections can be obtained by effective classical algorithms as well. 

But the quantum formalism of Hilbert spaces says that the exponentially small amplitudes $\la$, (which cannot be observed directly in any experiments because the time $1/|\la |^2$ needed for their detecting is too large) can interfere constructively in huge quantities and result in really observable values. In addition this interference can be organized so that it fits in an admissible time. This is the essense of the quantum computing. But we still do not know are there such effects in the reality or not. At least all the phenomena that have the theoretical substantiation do not require the existance of exponentially small amplitudes for this substantiation. 

The algorithmic description of physics can be thus in principle possible. If we assume that the success of approximate methods like Hartree-Fock in molecular computations is not accidental then it may be real to create a general effective algorithm for the simulation of many particle systems. Algorithmic description differs radically from the traditional because it is based on the notion of algorithms instead of analysis of infinitesimals; this new formalism can be called algorithmic physics\footnote{Mathematical analysis is the traditional sound basis on which the physical intuition is always brought up. Nevertheless, it is important to understand that everything has its limits. The standard analysis is thus good as the physical formalism up to the moment when it leads to something no computable. There exists the cut version of mathematical analysis where only computable functions are considered - the so-called constructive analysis. It radically differs from the standard analysis; for example, all functions there are continuous. This version of analysis in many respects better fits to physics than the standard one.
The method of amplitude quanta described in the Appendix 1 gives the description of quantum states just in terms of constructive analysis. But the method of amplitude quanta is the particular method, and the algorithmic approach cannot be reduced to the replacement of standard analysis by the constructive one. Particularly, the usage of discontinuous functions is sometimes very fruitful for the construction of algorithms. In general, in the algorithmic physics all descriptive tricks are admissible, but only in the framework of the effective computational procedures.}. I venture to call this approach the algorithmic physics; this name implies no analogies but only underlines the principal difference between this approach and the conventional understanding of physical theories. In its general form it can be considered now as a hypothesis, alternative to the hypothesis of a scalable QC. The single chance to refute this hypothesis is to build a QC, we see no other way to reject the algorithmic physics despite of that it is very different from the usual physics. 

\section{Main features of algorithmic physics}

\subsection{General notes}

Why do we need algorithms in physics and why the standard analysis and algebra are not sufficient? 
Strictly speaking the physics needs algorithms as an auxiliary tool that is designed for the solution of equations which express physical laws. Algorithms and computers are traditionally used in physics forcedly, their usage is connected with the well known fact that the systems of equations expressing the many body dynamics in general case have no analytic solution. In other words, if even we are able to express laws by formulas we cannot derive analytically practically important corollaries (trajectories) from them. Just this gap in the traditional formalism was filled by algorithms. In the algorithmic approach to the classical physics the derivatives are replaced by the corresponding difference schemes that reduce the problem to the tasks of linear algebra. The main drawback of this scheme is the instability of classical trajectories. Small perturbation in the initial conditions leads to the large divergence in the limited time that makes the method of finite differences in many cases inefficient. But this obstacle seems not fatal. The laws of classical dynamics lose its force when the distances becomes less than  $10^{-8}$ cm, because in this area we must use the laws of quantum mechanics. Quantum evolution of a particle is a unitary operator on its wave function that preserves distances, hence the small perturbation of the initial conditions can only lead to the small divergence of the trajectories independently of the evolution time (situation will be different if we admit measurements). At least in quantum mechanics we have no such an obstacle for the algorithmic approach as in the classical case. 

The algorithmic approach cannot influence to the part of physics that is already known, because all the known facts can be derived from the main laws and from some set of natural and simple assumptions by effective classical algorithms. The explicit difference from the traditional understanding of physics in this approach is connected to the possibility of the creation of a scalable QC which is allowed in the traditional physics and forbidden in the algorithmic physics. But this device still lies very far from the usual experiments and can be treated as a kind of abstraction. In addition one could suppose that if even a scalable QC is possible, its practical implementation is too far from our possibilities in the foreseeable future, and this argument for the reconciliation with the traditional viewpoint would seem valid. One could thus conclude that the algorithmic approach simply reflects the practical attitude of programmers to physics and we cannot wait from this approach more than improvements of the existing numerical methods.
But this opinion is wrong. Algorithmic approach differs radically from the traditional because it gives some new understanding of the physical problems and new treatment of such a key phenomenon as decoherence. The severe limitation of the classical computational limits for simulation dictates the "cut-off" description of unitary evolution as compared to hilbertology\footnote{This term is proposed by Sergei Molotkov.}. Such a description must contain soft measurements of a current state because we have no sufficient computational resources for the exact simulation of a unitary many body evolution, and these measurements are sufficient for the simulation of decoherence. We thus must not look for the decoherence specially, because it arises in the model independently of our wish, as the measure of the deviation of the classical description from Hilbert formalism for many particles. 

The most natural way of such a "cut-off" of Hilbert formalism is as follows. We simply do not account the deposit of states with too small amplitudes to the wave function. Namely, let $T$ be the accessible size of a computational resource (the number of steps of an algorithm or the number of elements in the memory). We then consider as zero all the amplitudes $\la$ such that $|\la |<\frac{1}{\sqrt{T}}$. It means that we consider as impossible such an event whose quantum probability is too small to make it observable in the accessible time frame $T$. Of course, we have no method to determine the factual value of $T$, but if we choose it starting from the capacity of the existing computers we could simulate many particle evoltions with the maximal account of all quantum effects. We thus assume that an amlitude is not continuous but grained, where its grain $\e$ - is the minimal nonzero value of an amplitude module is so small that its direct measurement is impossible because of the huge waiting time $1/\e^2$ so rare events. But if $\e$ is not exponentially small this must become apparent in the many body quantum problems, in particular it makes impossible the creation of a scalable quantum computer. 

We shall see below that this simple rule of cut-off easily gives two important things: the explanation of Born rule for the calculation of a quantum probability as a squared module of amplitude, and the uniform description of a quantum dynamics including unitary evolutions as well as a decoherence. Moreover, this rule makes possible to obtain the classical description of dynamics from the quantum description without any artificial tricks. The desirability of such a uniform description was expressed from the very beginning of the history of quantum mechanics\footnote{See, for example the famous polemics between Einstein and Bohr. In several recent works this discontent with this strange feature of quantum physics leads to the attempts to find its connection with the phenomena of consciousness (see, for example \cite{Pe}, \cite{Ha}).}. We produce some reasons for that such a description can be obtained in the framework of the algorithmic approach. 

Acceptance of an algorithm as a basic notion of physics instead of the analytical and algebraic formalism leads to the far-reaching consequences. Computational methods that give a good approximation to experiments must be accepted as a first principle description where any inaccuracy is regarded as a defect in used algorithms or as a bad initial data. The analytical formalism must be then considered as the form of instructions for the composing of simulating algorithms and the tool for debugging. The limitations of the purely algorithmic nature must be treated equally with the fundamental physical laws. 
Particularly it means that the irreversibility of quantum state corruption in a measurement or in a decoherence must be treated not as the result of the action of observer but as the result of shortage of the computational resource for the description of current state. This treatment is absolutely unacceptable from the traditional viewpoint but it leads to no explicit contradictions. This removes an observer from the description of quantum dynamics and gives to the algorithmic formalism such a completeness that is lacking in the standard quantum theory. 

The following feature of the algorithmic approach is that the model must be divided into two segments: the user segment and the administrative segment that is connected with the applicability of the "free will" principle. All the part where this principle is valid belongs to the user segment. The rest part of a model that contains the information to which this principle is inapplicable and which is only needed for the right "film" showing belongs to the administrative segment. 
For example, the coordinates of all the points in the considered area of the space-time belongs to the user segment because a user has free access to this area. Any trajectory in the light cone lies in the user segment because it can be realized in principle. In general, any process that can be described in terms of the so-called local realism (e.g., without the quantum long-range action) must belong to the user segment. The simplest explanation of the necessity of the administrative part is shown by entangled states of photons (EPR pairs).  
If two detectors measuring photons are disposed at the large distance one from another then we cannot simulate the detection of EPR pair by the user part only. 
Really, let us imagine that the orientation of one detector changes so fast that the light signal about this change cannot reach the second detector in the time of experiment. Having the "free will" in the user part to which the both detectors belong we can do it and randomly. The statistics of the second photon measurements must not then change in comparison with the case when the first detector is fixed, but the joint statistics will change. 
If we have the user segment only we cannot simulate this experiment without the assumption that some object transmitting a user's information moves along a trajectory which goes outside the light cone that is impossible. We thus see that the administrative part of the model is necessary for the right description of the quantum long-range action. 

The weak side of the algorithmic approach is that it can be contextual. If we limit our consideration by classical algorithms with polynomial complexity, then for the description of quantum systems we must somehow restrict the growth of the dimensionality of the space ${\cal H}$ of states, that means the choice of some subset ${\cal H}_0$. For example for the electronic configurations of a molecule when the spatial positions of its atoms nuclei are fixed, the choice of  ${\cal H}_0$ is reduced to the choice of one electron functions and their groups from which the Fock-Sleter determinants are formed. But the numerous works on quantum theory of molecules witness that there can be no universal way to choose these functions that is valid for all molecules. 
It means that the choice of basic wave functions for one particle may depend not only on its type but also on its vicinity, e.g., on the positions of the other particles (in the case of molecules it is the position of atoms). The properties of the particles essential for the algorithmic approach may depend on a content within which these particle are considered. 
If we speak about the simulation of the dynamics it means that the choice of subset ${\cal H}(t)_0$ depends on the state of environment ${\cal H}(t_{env})_{env}$ in the moments $t_{env}\leq t$, where $t$ denotes here not the physical time but the time in the administrative part, that is proportional to the number of steps of the simulating algorithm\footnote{The connection of this time with the local physical time depends on the simulated space, and we will touch this subject in the Appendix 2.}. This connection of the considered quantum system with the environment is determined by the entanglement and it is unavoidable in any approach to the simulation of quantum systems. 
Hence to describe the considered system and to choose ${\cal H}(t)_0$ optimally it is necessary to have some a-priori model of its behavior. The simulation in the usual sense when we fix the initial condition independently of the model and obtain the result at the end - may be sometimes impossible. If we do not know beforehand the form of
 ${\cal H}_0$, then the single way will be to consider the whole Hilbert space that immediately leads us to the insurmountable phenomenon of QC.

This difficulty results from the following evident fact. The size of computational resources that we have at our disposal can be roughly estimated by the number $10^9$. In the foreseable future this number can grow 3-4 orders at most, mainly due to the parallel computations and the creation of computational clusters. In the same time the number of atoms in the density packing in one cubic centimeter is about $10^{24}$. We see that the gap between our computational capacities and the sizes of systems that we plan to simulate is more than 15 orders. If we account the spatial degrees of freedom that is necessary for the real simulation even without entanglement, this gap will grow up to about 24 orders. Even if the main hypothesis of the algorithmic approach is right and the Nature is the gygantic computational net, the direct modeling of such a net by the known computational tricks that are factually "calculations by formulas", like the method of finite differences can be successful if only the growth of total numbers of atoms in the considered system does not lead to the substantial change in its behavior (for example, for regular crystalls). But in the very important cases when the complexity of the behavior essentially depends on the system size such methods lead to the straight competition of our computer with the real system and here we have no chance of success. Quantum effects, entanglement are among such cases. 

The existence of such effects is proved in the numerous different experiments but we know too little how they influence to the well known processes with observable results, for example to chemical reactions. Factually, almost all that can be taken from the analytic and algebraic approach is still embodied in the direct computational methods that we spoke about earlier, and moreover - in the existing software products. In what follows we can believe mainly in the computational procedures of the different kind that originate from the so called semi empirical methods. The genetic algorithms belong to this kind of procedures. The simulated many atom structure can be divided into the small parts and use the different methods of finding of the electron states in each part. After some time we can compare the results and choose from these methods the small number of those which give the most adequate picture of the evolution. We can then combine such selected methods and replace by them and by their combinations all the others. We then repeate this loop again and again varying the parameters of the selection depending on what part of the structure they are applied to, etc. As the methods we can use: the form of a probe function that approximates the exact wave function, or the methods of finding of the direction and frequency of emitted photons, or the form of amplitude quanta trajectories (see Appendix 1). The passage to such purely algorithmic constructions in the simulation seems unavoidable. 

In any case we must have a-priori representation about the behavior of the simulated system that will be specified after each user's review of the "film" based on this representation. The debugging of the model will have iterated character and in each step of it the user will have more and more exact picture of the simulated evolution. 
Just this process of debugging will replace the axiomatic building of quantum physics (see (\cite{BS}). 
Of course, it is not a breaking-off with the tradition quantum theory but the change of accents only. For example, the conventional formalism of quantum physics will be not the main instrument but rather the tool for debugging of simulating programs. 
 
The aim of this article is the discussion of some possible ways of the development of algorithmic physics independently of the general fate of this approach. This discussion can be useful to those who try to simulate the processes where quantum effects play the substantial role.

\subsection{User and administrative parts of a model}

It follows from the above explanation that the principal difference between the algorithmic and standard approaches results from the nature of algorithms: in general case there is no method to learn the result of their work on a given initial data but the sequential fulfilling of elementary steps determined by this algorithm\footnote{This thesis remains valid even if we use a quantum computer for the prediction of the work of classical one. The majority of not long classical computations do not allow quantum speedup even on one step (see (\cite{Oz}).}.
Some corollaries from this surprising feature are discussed in the next section. Here we consider in more details the general structure of the algorithmic model, that has been already mentioned, namely its separation into the user and administrative segments that can be treated as the peculiar discretionary access control.

Since the model must show the dynamical picture of the system behavior its user segment must contain the description of objects with the physical sense - elementary particles. The administrative segment consists of elements with no physical sense. The necessity of the administrative part is substantiated by the known experiments (see, for example, (\cite{As}, \cite{B}, \cite{Be})), establishing the impossibility of the local realism in quantum physics or, in other words, the violation of the Bell inequalities. 

The simplest example showing the necessity of the administrative part in the model for the massive particles is the demonstration of its entangled states. We consider the pure state of the system of the two particles of EPR type:
\begin{equation}
\Psi=\a |0_10_2\rangle +\b |1_11_2\rangle ,
\end{equation}
and try to distinguish it from the mixed state $\rho$, in which the fractions of the pairs in state $|0_10_2\rangle$ and $|1_11_2\rangle$ are $|\a |^2$ and $|\b |^2$ correspondingly. To attach the physical sense to this situation we assume that $|0_j\rangle$ and $|1_j\rangle$ denote the spatial positions of the particle $j$, $j=1,2$. Intuitive sense of entanglement of the state $\Psi$ is that not only the coordinates of the two particles are strictly connected (here simply equal), but also impulses; this is just the difference of the state $\Psi$ from the mixture $\rho$. If we measure only the coordinates of both particles in state $\Psi$, we obtain exactly the same result that will be if the system is in state $\rho$, so this measurement cannot distinguish these two cases. But if we measure impulses of these particles we find out the difference between $\Psi$ and $\rho$. If in the first case impulses will be always equal, then in the second case we obtain the full dispersion in the measured values due to the uncertainty principle applied separately to each particle which are independent in state $\rho$ and which can be thus considered as patterns of the same particle that is in the state $|0\rangle$, or in the state $|1\rangle$. For the substantiation we must turn to the impulse representation of the wave function. In the chosen designations the Fourier transform giving the impulse representation of the wave function can be replaced by its zero approximation: Hadamard transform of the form 
\begin{equation}
\left(
\begin{array}{lll}
&\frac{1}{\sqrt{2}}&\frac{1}{\sqrt{2}}\\
&\frac{1}{\sqrt{2}}&-\frac{1}{\sqrt{2}}
\end{array}
\right),
\end{equation}
that is applied to each qubit. It is straightforwardly verified that it remains the state $Psi$ unchanged and the following measurement (it will be the measurement of impulses) again gives the equal values for both qubits. 
In case of mixture $\rho$ the situation will be different. Here Hadamard transform applied to each qubit gives the mixed state in which there are the both pure states $\frac{1}{2}( |00\rangle + |01\rangle + |10\rangle + |11\rangle )$ and $\frac{1}{2}( |00\rangle - |01\rangle - |10\rangle + |11\rangle )$ with probability $1/2$ each. Hence in the case of initial state $\rho$ the measurement of impulses gives the uniform distribution of impulses of the both particles among all possible combinations. One can checked that this conclusion is also valid if we take the first approximation of Fourier transform (with the $\pi /2$ phase shifts) instead of zero approximation. We see that there exist entangled states principally different from mixtures of not entangled states and such states can be detected in experiments not only with photons but with the massive particles as well.

We conclude that it is impossible to simulate a wave function collapse and entangled states basing only on the local interactions, and it is necessary to have the administrative segment of the model. The administrative segment is not accessible to users; in particular they cannot obtain its state in a given moment. It contains the data that is called hidden parameters, but it is important that these parameters are not local - they are connected with the spatially distant points. The information determining the entanglement is stored just in this segment of the model. Users cannot address directly to this administrative channels to the information exchange. The simulation of moving particles has its special restrictions. We cannot simulate an arbitrary speed of the movement because it is necessary to trace the passing of a particle through all intermediate nodes of the spatial grid used in the simulation, and each such node requires some amount of the time. There are thus the limits on the possible speed that can be simulated and this limit is determined by the frequency of the simulating processor. We thus see that the relativistic limit on the information transfer results from the inaccessibility of the administrative part for users\footnote{One of the possible ways for the simulation of the relativistic pseudo Euclidean metric in the space-time is shown in the Appendix 2.}. 

We treated the division of the model into two segments as the peculiar discretionary access control. For the simulation it is necessary that the administrative segment, at first disposes the complete information about the intentions of a user, and secondly has the possibility of instant access to the remote points of the physical space.  We nevertheless assume that the possibilities of the administrative part are not boundless but limited in its turn by the theory of classical algorithms. In particular, the memory accessible to the model grows linearly with the size of simulated physical space. It can be formalized by a multihead Turing machine, for which the instant access means the application of rules containing states of many heads. It is described in more details in Appendix 2. We note here the connection between the discretionary access control and the applicability of a "free will" principle. The priority of the administrative segment in our model factually means that a user's "free will" is conditioned explicitly on its connections with the unlimited external world\footnote{By external world we mean not only the macroscopic and megascopic universe, but the potential microscopic universe as well, e.g., the possible structure of elementary particles.}. If there are not such connections a user itself could be included in the model with the sufficient computational resources. 

\subsection{Description of measurements. Obtaining of Born rule for quantum probability}

Born's rule for the quantum probability has the form 
\begin{equation}
p(A)=|\langle A|\Psi\rangle |^2,
\label{p}
\end{equation}
where $A$ is a vector belonging to a basis $e_1,e_2,\ldots$ corresponding to the measurement, $|\Psi\rangle$ - is a measured state. (In the physical terminology $e_1,e_2,\ldots$ is a basis consisting of eigenvectors of Hermitian $H$ which determins this measurement.) Born rule is the single link of the traditional (copenhagen) formalism that connects quantum mechanics with the classical and this rule is assumed as a key axiom in this formalism. The status of this rule makes impossible to obtain a unified description of quantum dynamics which would be independent of the existence of an observer (factually of an observer's "free will"). This is why attempts to derive Born rules from something more fundamental do not end up to nowadays. The central point and the main cause of failure in this direction was the absence of a classical urn scheme for the quantum probability that would reduce Born rule to the frequency definition of probability, and that would be natural from the physical viewpoint (an artificial introduction of an urn scheme is possible but it is not interesting). One of the last attempts was done by Zurek (see. (\cite{Zu})). His proposal is based on the operation of swap between quantum states leading to the equality of amplitude modules of elementary events, that is not completely natural from the physical viewpoint. (see (\cite{Mo}, \cite{SF})). That proposal is based on the standard approach with Hilbert states in the spirit of Gleason's theorem ((\cite{Gl}), see also (\cite{CFS}, \cite{Bu})\footnote{Already after the finishing of this paper my attention to the Zurek's interpretation and to the series of articles connected with it and with Gleason's theorem was attracted by A. Sheverev (\cite{She}). This theorem says that every nonnegative function on vectors in a Hilbert space of dimensionality more than 2, which is a probability measure on all basices of this space has the form (\ref{p}) for some vector $\Psi$. The limitation on the dimensionality is the indirect evidence of the redundancy of this theorem for the quantum physics, because in the reality we always deal with some concrete wave function (for the dimensionality 2 the counterexample is straightforward).}. The description of Born's rule represented below differs from Zurek's in that it is based on an amlitude quantum but not on the swap operation.

We give the description of Born's rule starting from the concept of an amplitude quantum.

The consideration of quantum evolution from the viewpoint of the many particle Hilbert formalism gives the states of the form

\begin{equation}
|\Psi\rangle=\sum\limits_j\la_j|e_j\rangle,
\end{equation}

where the summing is spread to the infinite set of basic states $|e_j\rangle$. The algorithmic approach requires the constriction of this sum to the finite sum by the cutting of all the summands with coefficients $\la_j$, which modules are less than some fixed threshold $\e$. Such a sum will contain no more than $1/\e^2$ summands. Let $N$ be the number of the basic states for one particle. We can take $\e=\frac{1}{\sqrt{N}}$. The resulting state thus has the form

\begin{equation}
|\Psi\rangle=\sum\limits_{j=1}^N\la_j|e_j\rangle,
\label{limstate}
\end{equation}
where some summands can be zero. 

This procedure of elimination of all summands which modules of amplitudes are less than $\e$ is called a reduction. This constant $\e >0$ is called an amplitude quantum. 
We agree to fulfil a reduction over each state that we obtain in our simulation process. Such reduced states are called admissible.

We now show how the reduction, e.g., nulling of the small amplitudes, gives Born rule for the finding of quantum probability. Our aim is to reduce the finding of probability to obtain a certain basic state $A$ in the measurement of a quantum state $\Psi$ to the application of the classical rule 
$$
p(A)=\frac{N_{suc}}{N_{tot}}
$$
where $N_{suc}$ is the number of successful outcomes (e.g., such elementary events which mean the realization of the event $A$), $N_{tot}$ - the total number of all elementary events. We have to define the set of all elementary events and establish the correspondence between them and basic states of the system. We call elementary events such basic states of the extended system (measured system + measuring device) which amplitude modules in a given state equal to an amplitude quantum $\e$. A set of elementary events thus depends on a quantum state of extended system. 

Let $|\Psi_j\rangle$ denote basic states of a considered system and $|\Phi_j\rangle$ denote basic states of a measuring apparatus (that can be an eye of observer).  The contact between these two objects results in the state of the form
\begin{equation}
\sum\limits_j\la_j|\Psi_j\rangle\bigotimes |\Phi_j\rangle
\label{meas}
\end{equation}

Since the measuring apparatus is very massive in comparizon with the measured object, when trying to describe its quantum states we have to split the states from (\ref{meas}) to the sums of $l_j$ basic states (all states of numerous particles inside the measuring apparatus must be taken into account, like nuclea, electrons, etc.). In the other words, even if in the instant of contact there was the state  $|\Phi_j\rangle$, the evolution very quickly transforms it to a state of the form $|\Phi'_j\rangle =\sum\limits_{k=1}^{l_j}\mu_{j,k}|\phi_{j,k}\rangle$, where all $l_j$ grow very quickly up to the instant when amplitudes reach the value of an amplitude quantum and they will be nulled. Hence all the modules of amplitudes $\mu_{j,k}$ must be then taken as approximately equal. If we substitute the expression for $|\Phi'_j\rangle$ instead of $|\Phi_j\rangle$ into the (\ref{meas}), the amplitude of states $\phi_{j,k}$ will be about $\frac{\la_j}{\sqrt{l_j}}$ due to the unitarity of evolution. 

We have to fulfil the reduction that is to null all summands $\phi_{j,k}$ which amplitude is too small. Since the time frame when the splitting to such summands happens is negligible, in the computations it means that we split each summand in (\ref{meas}) to $l_j$ new summands so that modules of all resulted amplitudes are close to the amplitude quantum and approximately equal, because only this supposition makes this splitting equitable to all the states before the reduction that is required for the implementation of a classical urn scheme\footnote{Approximate equality of amplitude modules before the reduction corresponds also to the urn scheme based on amplitude quanta; see, for example, Appendix 1.}. But the total number $l_j$ of the summands with the first factor $|\Psi_j\rangle$ is exactly the total number of successful outcomes, and it is proportional to $|\la_j|^2$, and if exactly one of them survives in the reduction we obtain the Born rule for the quantum probability. 

A probability space thus depends on the choice of a wave function $|\Psi\rangle$. We consider factually the conditional probabilities to obtain this or that result in the measurement of the system provided it is initially in a state  $|\Psi\rangle$. We note that despite of the apparent narrowing of formulation comparatively to Gleason's theorem just such probability spaces have the physical sense. 

This explanation of Born rule is based on the notion of reduction of quantum state as the nulling of too small amplitudes. We agree to fulfil this reduction at each step of the simulation of a quantum evolution because otherwise the simulation would be impossible at all. In our approach the specificity of a measurement comparatively with the unitary evolution is only quantitative: a measurement happens in an instant when the system comes into contact with the massive object that can be called an environment. It results in the splitting of the sumnads in (\ref{meas}) to the big number of new summands. In addition to this natural supposition we used only the norming of the wave function which conservation results from Shroedinger equation. In the explanation of Born rule no suppositions were applied that exceed the bounds of conventional agreements of quantum mechanics but one: the reduction of a wave function that is treated as a nulling of small amplitudes. Just this procedure of reduction transforms the set of Feynman paths to the classical trajectory in case of a massive body (see Appendix 1). We treat the decoherence as the forming of entangled states of the (\ref{meas}) with the environment e.g., we do not distinguish it from the measurement of our system. Born law for the quantum probability and irreversible corruption of a state resulted from the decoherence thus follows from the grain of amplitudes. 

Algorithmic approach thus gives the unified description of a unitary evolution and a measurement that gives the independent of an observer description of quantum dymanics. This is the advantage of algorithmic approach, because the conventional formalism does not give such a description and depends principally on the presence of observer\footnote{That immediately raises the question about what object posesses the status of observer. All who considered the foundations of quantum theory noticed this paradoxial situation. It is important that such a dependence of quantum theory on an unintelligible object which causes the decoherence (and the unclear treatment of the decoherence itself) deprived quantum theory of the possibility to submit all area of molecular phenomena, and first of all such phenomena that are crucial for the functioning of living organisms.}. 

In the algorithmic simulation we thus must not especially account that somebody observes our system. Moreover, an observer itself (if any) can be included to the simulated system without any change of the simulating algorithm, provided this observer is independent of environment. The single reason that makes impossible to simulate itself (that would lead to logical contradictions) is insurmountable limitation of the somputational resources, because for the exact simulation of some system the other system is required which is much bigger.  

\subsection{Hierarchical model of quantum many particle dynamics}

The central point of the algorithmic approach is the choice of subset of states $\cal{H}_0$ of the simulated system which description must grow not too fast when the number of particles increases. The admissible speed of growth is the linear because only in this case we have at least a theoretical possibility to create in future (with the most powerful of classical computers) "films" describing the behavior of the living matter\footnote{Even in the case of quadratic speed this hope will disappear.}. The choice of such a subset $\cal{H}_0$ is the radical break with the many particle Hilbert formalism and with the hope to simulate a scalable quantum computer. This subset in the general case is not a subspace because we base not on analytical properties as linearity but on a possibility to describe an evolution by effective algorithms. From the traditional viewpoint it means that we choose the approximation to the solution of many particle Shroedinger equations. There is too big uncertainty here to use purely algorithmic heuristic tricks like genetic algorithms, and we have to explicitly point the form of this subset. We start with the evident for the algorithmic approach limitation on the complexity of the quantum state notations for $n$ particle system, that is the length of this notation in terms of sums and tensor products must be limited from above by some constant\footnote{See (\cite{Aa}).}. It is naturally to assume that this constant must depend linearly from $n$. We show how the quantum dynamics can be simulated in terms of such states. We represent some heuristic arguments for it that issue from the feature of algorithmic approach and partially generalize the known set of computational tricks used in quantum calculations. The main of these arguments is that the method proposed is closest to the one particle description in sense of the algebraic notation of states.

The direct method of simulation looks as follows. We consider not an evolution of a wave function $|\Psi (t)\rangle$, but an evolution of a pair of the form $|\Psi (t)\rangle ,\ P(t)$, where $P=\{ \bar x_1,\bar x_2,\ldots, \bar x_L\}$ is a set of points of division of the configuration space for many particles such that their density is proportional to the squared module of the wave function: $\rho (t)\equiv \langle\Psi (t)\ |\ \Psi (t)\rangle$, and their total number $L=\frac{1}{g^2}$, where $g$ is a given value of the amplitude quantum (that is mush less than the value existing in Nature). For the simplicity we can assume that these points are located so that the difference scheme for Laplacian for each of the particles in the considered system is applicable with them. Moreover, we can assume that that for every fixation of any $s-1$ particles the total number of the points of division for the rest one and their density obey the same law. The following wave vector $|\Psi (t+\Delta t)\rangle$ then is obtained from $|\Psi (t)\rangle$ by the application of the dinite difference scheme for Shoedinger equation and the new set of the points of division $P(t+\Delta t)$ is obtained from the new wave function accordingly to the condition of density stated above. For the improvement of the values of the new wave function in intermediate points which are included in the new division we can use the methods of approximation (for example, splains). Such a method is based immediately on an amplitude quantum value and it allows to account the entanglement of every type between the simulated particles in the framework dictated by this value. But just due to it this method can be not effective, because the real value of $g$ can be much less than that is addmissible to the real superconputers. We then describe the approach which is more universal for the computations. It is based on the conception of the hierarchy of particles and the direct method will work as a part of it in the moment when this hierarchy will be rebuilding. 

 Since our aim is to learn how to create realistic "films", all our approach must be based on the concept of particles which will be the main objects of such "films". 
For the scalability we must keep in mind that every particle (perhaps, but photons only) can consist of more elementary particles and our approach must admit the corresponding scalability. 
We thus consider some groups of particles as particles as well; for example nuclei, atoms and molecules, or more special groups as Cooper electron pairs and quasi particles, e.g., all the groups which can be considered as a whole particles. Here we consider a group as a particle if the application of Shroedinger equation for one particle (that is the single type of Shoedinger equations which can be solved by classical computers) gives the sensible result for this object. We further concretize this informal explanation. We separate the class of maximal particles considered in the model - they will have the zero level, whereas the elementary particles which cannot be splitted by using the considered interactions will have the biggest level. We restrict our consideration by electromagnetic interactions, thus in our case electrons, photons and nuclei will have the biggest level\footnote{Nevertheless, the proposed approach is seemingly applicable to the nuclear interactions as well. At least, the possibility of ranging in the line to the increasing of embedding depth of particles is exists in the hierarchical model. It is convenient for the unified description of the different types of interactions, for example, electromagnetic and nuclear.}.

We consider two main computational tasks: the simulation of unitary evolution, e.g., the modeling of operator  $\exp(-\frac{iH}{h}t)$, and the finding of eigenstates $|\phi_k\rangle$ of some Hamiltonian $H$. Let us estimate the time required for it if we use the direct method. For simplicity we consider a system with 2 particles. For such a system the total number of basic states is $N^2$. The matrix of Hamiltonian has the dimension $N^2\times N^2$ and one step of the evolution requires $N^4$ elementary operations, hence for the time frame $t$ the total number of them cannot be less than $N^2t$, that takes place for the method of finite differences applied to Shroedinger equation. For the task of finding eigenstates we have to solve the characteristic equation for the matrix of the size of the order $N^2\times N^2$, that requires about $N^{12}$ operations. If we use the space grid with $10$ points to each dimension that is the less admissible accuracy for one particle we have $N=1000$ and the finding of eigenstates of two particle Hamiltonian requires $10^{36}$ operations that makes the direct method useless even for supercomputers and particular tasks where the Hamiltonian has symmetric form. In the practical computations for such tasks the conventional methods are: the method of density functionals (see (\cite{LA}), or Hartree-Fock method. Hartree-Fock method is based on the representation of many particle wave function for a system of $n$ identical fermions as a Fock-Sleter determinant (see (\cite{Sl})). 

It means that we account the entanglement between particles that comes from their exchange interaction but not from Coulomb interaction. To account the entanglement that comes from Coulomb interaction we should represent the wave function as a sum of determinants of the form (\ref{sym}):

\begin{equation}
\sum\limits_j\mu_j|\Psi_j\rangle .
\label{gen}
\end{equation}

But this representation contains the infinite row and the direct generalization of Hartree-Fock method to this case gives a boundless problem instead of  robust method for (\ref{sym}) because we have no guidelines for the choise of $\mu_j$, for  example, if some of them is not negligible in comparizon with $\e$, we cannot account the corresponding summand. The density functional method does not account the entanglement at all; it is fine for the cases where the density of wave functions is almost the same (for example, for electrons in metals), but for atomic and molecular computations this method gives a big error.   

We now describe the method that makes possible to account all the types of entanglement between the particles in the assumption that the real amplitude quantum $g$ is much less than the value of $\e$ which is equal to $\frac{1}{\sqrt{T}}$, where $T$ is the total number of elementary operations of the fastest real supercomputer in the maximal time frame in our disposition. 

The maximal particles are called the particles of the zero level. 
The particles of the first level will be the biggest parts of particles of the zero level, etc. The choice of particles of a level $n$ thus means the choice of the grouping of particles of the level $n+1$; in the first step this is the task for a user. For the further steps we will formulate the rule for the change of this hierarchy. The general recommendation is only that this procedure must give the objects to which the application of the notion of wave functions and Shroedinger equation leads to the sensible result. 
Each particle $a$ of the level $n$ thus has its spatial coordinates $x_a,\ y_a,\ z_a$ and spin coordinate $s_a$. These coordinates can be often treated as coordinates of the centre of mass $C_a$ of the set of minimal particles forming $a$ and all particles inside of $a$. Let $a_1,a_2,\ldots,a_s$ be the particles of level $n+1$ that form $a$. Their coordinates in the coordinate system which initial point is $C_a$ are called the relative coordinates. 

In what follows we will use the qubit notation of wave functions $|\Psi(\bar r)\rangle$ in the form
\begin{equation}
\sum\limits_{\bar r}\la_{\bar r}|\bar r\rangle
\label{wave_function}
\end{equation}
where $\bar r$ is a binary notation of numerical value of coordinates of all particles in the considered system; let the length of this string be $n$. Here a value of an ordinary wave function $|\Psi(\bar r)\rangle$ is proportional to $\la_{\bar r}$. We assume the natural lexicographic order on the string $\bar r$ which exactly corresponds to the case of one particle in one dimension space, but our consideration will be general.\footnote{ The representation of wave functions in the form (\ref{wave_function}) is much more convenient than in the traditional for physicists form $|\Psi(\bar r)\rangle$, because the last form is ambiguous, it means two different things: the wave function and its value in a concrete point $\bar r$ (so that to tell apart these two senses physicists often write integrals with delta-functions).} Since we agree that any particle of a level $k-1$ is located in the center of mass of the particles of level $k$ that form it, in this group of particles a fixation of all but one particle determins a coordinate of this one (relatively to their center of mass). These particles which coordinates can be arbitrary, are called valuable. Let $k=0,1,\ldots ,n$ enumerate the levels of hierarchy. We denote by $\bar r_k$ the initial segment of sequence $\bar r$ of the length $k$, and by $r_k$ - $k$-th element of this sequence, that has the form of list $r_k=(r_k^1,r_k^2,\ldots,r_k^{s_k})$, where $r_k^j$ are the relative coordinates of $j$-th valuable particle of a level $k$, $s_k$ is the total number of such particles; for example, if any particle contains exactly two particles of the next level, then $s_k=2^{n-k-1}$. If the upper indices are not used we can assume for simplicity that $r_k$ is the single qubit - this simplifies our notations. Each wave function of the form (\ref{wave_function}) can be represented as
\begin{equation}
\sum\limits_{r_1}\left(\la_{\bar r_1}|r_1\rangle\bigotimes\sum\limits_{r_2}\left(\la_{\bar r_2}|r_2\rangle\bigotimes\ldots\bigotimes\sum\limits_{r_n}\la_{\bar r_n}|\bar r_n\rangle\right)\ldots\right)
\label{hie}
\end{equation}
For this it is sufficient, for example, to take all $\la_{\bar r_j}$ equal to 1 for $j=1,2,\ldots,n-1$, and for $j=n$ to set it equal to $\la_{\bar r}$ from the formula (\ref{wave_function}). 

If we fix some value of $j$, the amplitude distribution $\la_{\bar r_j}$ can be treated as some wave function; we assume that it is normed. Accordingly to our agreement it can be determined by some effective algorithm $f_j$, which code is denoted by $[\bar \la_{\bar r_j}]$, so that $f_j(\bar r_j)=\la_{\bar r_j}$. Let $K_j$ denote the set of lists of the form $\bar r_j$, and let $F_j$ be such a function on $K_{j-1}$, that $F_j(\bar r_{j-1})=[f_j]$. We will consider such states only for which all the functions $F_j$ $j=1,2,\ldots,n$ can be computed by some finite and fixed set ${\cal A}$ of effective algorithms. Since a fixation of all functions $F_j$ uniquelly determines a state, all such states will be determined by a finite set of effective algorithms, where the length of codes of such states will be limited from above by some linear function of $n$, e.g. of the number of particles in the considered system. We note that in view of the last remark the states separated by (\ref{hie}) and your agreements represent the narrow subclass of all states (with the agreement about amplitude quantum). But the computations with the described class of states does not require the immediate storage of amplitude quantum in the memory; we have to store the codes of algorithms instead, that generate amplitude distributions - it makes possible to work with much less amplitude quantum than is allowed by the memory. 

Several subclasses can be introduced by the imposition of additional conditions. If all the functions $F_j$ depend factually not of the whole list $\bar r_{j-1}$, but of coordinates $r_{j-p},r_{j-p+1},\ldots,r_{j-1}$ only, we call such states the states of depth $p$. The subclass of states of the depth $0$ consists of non-entangleg states. If each distribution $\la_{\bar r_j}$ contains only one nonzero element, we obtain the set of basic states. 

We denote by $|\Psi_{\bar r_k}\rangle$ a wave function $\sum\limits_{r_k}\la_{\bar r_k}|\bar r_k\rangle$, which obviously depends on a choise of $\bar r_{k-1}$. It is a wave function of the system of all particles of level $k$, that depends on a choise of coordinates of particles belonging to the lower levels, enveloping particles of level $k$. We now treat this dependence in more details. Let $A$ be Hermitian in the space of states of a system $S_k$ paricles of level $k$. Its mean value is thus determined accordingly to the quantum rule 
\begin{equation}
\langle A\rangle_{\Psi_{\bar r_k}}=Tr\ (A|\Psi_{\bar r_k}\rangle\langle\Psi_{\bar r_k}|).
\end{equation}
In particular we can find the mean value of every component of the system $S_k$, and the potential $V_k(r)$, created by this system in a point $r$. Given an external potential $V$, we can find the potential $V'(\bar r_{k-1})=V+V_k(r)$, which acts on a particle of level $k-1$. If we are given initially eigenstates for particles of level $n$ we can thus compose the Hamiltonian for particles of level $n-1$; then find its eigenstates and thus compose the Hamiltonian for particles of level $n-2$, etc., up to the biggest particles of level $0$. An absolute coordinates of a level $k$ can be obtained as a sum of sequentially nested particles up to the level $0$. It turns out that a spatial fixation of particles of levels $k-1,k-2,\ldots,0$ determins an amplitude distribution for level $k$, that is required in the definition.  

A step of unitary evolution can be thus realized for a state of hierarchical system by some numerical method, for example, by finite differences. For particles belonging to the same tier we thus apply the method of direct simulation.  It is important that operations performed over amplitude distributions in this modeling and the resulting distributions lie in the set ${\cal A}$ of chosen effective algorithms. 

If we limit the total number of points in the space by a value $L$ (or fix a spatial grain), the quantity of eigenfunctions of every level and the maximal number of particles in each set $S_j$, then the memory required for the storage of any state of the form (\ref{hie}), will grow as a polynomial of the number $N$ of particles of the biggest level (elementary particles) and the degree depends on $L$. The hierarchical representation of wave functions given by the formula  (\ref{hie}), is not then equivalent to the many body Hilbert formalism, where the growth must be exponential. Newertheless the hierarchical representation of many body states gives the principal possibility to scale the quantum simulation not only for the systems consisting of elementary particles (atoms, molecules), but also inside elementary partices. 

We now describe how the defined hierarchy is varying in the time.  

1.) Lowering of a particle to one step in the hierarchy. We suppose that for the states of the form (\ref{hie}) the simulation of unitary evolution (with the mandatory reductions) leads to that a state of some particle of a level $k$ in each of functions $|\Psi_{\bar r_{k-1}}\rangle$ is separated as a tensor multiplier. We then declare this particle to belong to the level $k-1$ with the corresponding rebuilding of amplitude distributions. This particle will then interact with the other particles of the level $k-1$ accordingly to the corresponding Hamiltonian. 

We thus can describe the tearing off electrones from a molecule resulting from the Coulumb attraction of a close ion or an absorbed photon. Given initially two electrones in a Fock-Sleter state we consider the situation when the simulation of unitary dynamics with reductions leads to the growth of the distance between their one particle wave functions. The determinant then turns to one summand and we have the described situation.
The situation with photon absorbtion can be considered analogously. Here we must treat the states of the form
\begin{equation}
\sum\limits_j\la_j|\Psi_j,f_j\rangle ,
\end{equation}
where $f_j\in\{\ { photon\ in\ state\ }\psi_j,\ {no\ photons} \}$ (see below).

2.) Lifting of a particle $a$ to one step in the hierarchy. This process is reverse to the previous and it is connected with the creation of new entanglement between particles which were not entangled before. A particle $a$ then is included to the tier subordinate to one particle - $b$ with which $a$ was in the same level before. The criterion determining the moment for such a procedure is as follows. During the simulation of the dynamics of system of two particles $a$ and $b$ as a system of two interacting particles its state becomes entangled within the precision of simulation, and this entanglement does not dissappear after few steps. This criterion requires the direct many particle simulation. If we want to manage with one particle simulation only we can use the different criterion:

K). In the simulaton of system consisting of independent classical parts $a$ and $b$ it comes in that a fixation of coordinates and impulse of one determins the coordinates and impulse of the other within the accuracy of simulation. 

This rebuilding of hierarchy is the most nontrivial operation in the simulation, because it establishs the entanglement between particles which were independent before. The change of hierarchy represents the computational trick because the entanglement that arised initially in the immediate simulation of many body system (quantum or classical) turns to the hierarchical entanglement after the placing of initial point of the new coordinate system to the center of mass of a previously non-
entangled system.  

\nnn

{\bf Remark}. We could introduce the special procedure of measurement which is performed in the moment when the carrier of wave function of some particle becomes disconnected, e.g., comes apart to several components of connectivity $D_1,D_2,\ldots,D_k$. The measurement then would be the projection of wave function to one of these areas accordingly to the Born rule. But such a procedure in contrast to the reduction does not correspond to any computational principle and cannot be associated with any real process; the value of such a procedure would be purely  aesthetic, because it preserves the connectivity of the wave function carrier (that has indirect relation with the economy of the computational resources). For the disintegration to the different connectivity components (which can be far one from the other) the general description of a measurement procedure is applicable. This description is based on the reduction only and does not need any additional suppositions. This is because we do not introduce the special procedure of measurement. 

\nnn

We then define the division of the configuration space for particles in the hierarchy, that is needed for the numerical methods. If the points of the division are distributed uniformly it would result in a lot of redundant work, because the majority of basic states would have amplitudes which modules less than $\e$ and the corresponding summands will dissappear in the next reduction. In the area of big amplitudes the points of division must be disposed more densely because just these areas more influence on the evolution. In the passage from the conventional for analysis the discrete representation of continuous functions through the division of an interval by points $x_1,\ x_2,\ \ldots,\ x_k$ to the qubit representation (\ref{limstate}) we must choose these points so that the impact of the particle in one of the intervals of division corresponds to the basic state in the linear combination. It can be reached if we use a non-uniform distribution of the division points. 
How the density of the division points must depend on a wave function to minimize the computational resource required for the simulation of unitary evolution? If we start from the classical urn model for the quantum probability (see below) we should dispose the points so that they correspond to the elementary events. Namely, let $\rho (x)$ be the density of division points for configuration space. If $\la (x)$ is the wave function in its continuous representation the following condition must be fulfilled:
$\rho (x)=C\ |\Psi (x)|^2$ with some constant $C$. It guarantees the conservation of the wave function norm during the simulation. This trick with the non-uniform density of division points gives the best accordance with the procedure of state vector reduction when we ignore the small amplitudes. The idea of non-uniform density of the division points can be generalized to the hierarchical representation of a many particle system. 

For the simplicity we consider the case of two particles of the level 2 that form a particle of the level 1. (The generalization to the case of many particles is straightforward.) The points of division of configuration space for the 1 level particle are distributed acccordingly to our agreement about the density and their total number is
$[1/\e^2 ]$. If $x$ is the point of division for the 1 level particle that corresponds to the amplitude $\la$, then the total number of division points for one particle of the level 2 is $\left[\frac{\la}{\e}\right]$. The quantum evolution is simulated by the iteration of two steps: a)  one step of the evolution of 2 level particles when the 1 level particle is fixed, and
b) one step of the evolution of 1 level particle where the state of 2 level particles is fixed (in its coordinate system).
The simulation then requires the same total number of steps as with the uniform distribution of the division points but in the areas of bigger module of amplitude these points are distributed more densely that better corresponds to the ideology of simulation than the uniform distribution. 

For the description of ensembles of identical particles of high levels of nesting (for example, electrons) the representation in terms of eigenstates of energy is much more convenient than the language of coordinates, because such particles emit photons that change their states. It does not change the geheral scheme of hierarchical description, only eigenstates of the corresponding Hamiltomian are associated with the whole tier and by basic states $|\bar r\rangle$ we mean not a spatial positions but eigenstates of Hamiltonians. 
For the determining of the absolute coordinates of particles belonging to high levels of nesting we must, of course, pass to the coordinate representations of wave functions, though absolute coordinates can be hardly needed for anything. 

The finding of eigenstates requires the direct simulation that we will now consider. The starting point is that eigenstates $\Psi$ satisfy the following equation 
\begin{equation} 
\frac{\delta}{\delta \Psi }E(\Psi )=0, \ \ E(\Psi )=\int\Psi (r)^*H\Psi (r)\ dr
\end{equation}
This is the equation in variations of the wave function $\Psi$ is equivalent to the system of ordinary equations of the form
\begin{equation}
\frac{\partial}{\partial \la_j}E(\Psi )=0
\label{syst}
\end{equation}
for each $j$, where the wave function $\Psi$ is considered as the function of $\la_j$. Practically the system (\ref{syst}) can be solved by the sequence of steps. On each of them we choose the direction of the most increasing of the function $E(\Psi )$. The realization of each step requires the total number of operations proportional to the total number $M$ of division points of the common configuration space, where $M=N^k$, $N$ is the total number of division points of one particle space, $k$ - the number of particles. The total number of steps has the order $N^{\frac{1}{3}}$, that gives the total number of operations of the order $N^{k+\frac{1}{3}}$. For a many body system an initial wave function is typically chosen in the form of a tensor product of one particle wave functions: 
\begin{equation}
\Psi (\bar r_1,\ldots,\bar r_k)=\Psi_{i_1} (\bar r_1)\Psi_{i_2} (\bar r_2)\ldots\Psi_{i_k} (\bar r_k).
\label{prod}
\end{equation}
The described method of the minimization of energy must be at first applied under the condition that the general wave function has the form (\ref{prod}). It means that we vary functions $\Psi_{i_j}$, finding the minimal energy. As we find the set of one particle wave functions that give the minimal energy, we turn to the qubit representation of wave function (e.g., to the form (\ref{hie})), and then continue the energy minimization moving to the entangled states. 

The direct simulation for the particles of the same tier represents some difficulty if we cannot introduce the hierarchical order on them, as in the case of many electron states in atoms or molecular structures. The algorithmic realization of the direct method requires the extremely large resources consumption\footnote{Many dimensional grids of varying density can be easily built in the case of not entangled states only. For the entanglement of the general form the building of such grids is difficult.}, hence we now describe one trick that can valuably simplify the simulation. 

The idea of this trick is to account in the minimization of energy not all variations of wave functions but only such that corresponds to the basic states with the sufficiently large amplitudes. Here we will store wave functions in the form maximally close to (\ref{prod}). We will deal with the representations of wave functions in the form of formulas, and assume that the storage and the operation over these functions are fulfilled accordingly to such formulas. We aggree, that in tensor products the one particle wave functions are enumerated in the fixed order, and in the qubit representation of every one qubit wave function as a sum on all values of coordinates these values are chosen in the fixed order as well (for example, in the lexicographic). We shall not separate the spin coordinates from the spatial coordinates. 
Given a function $\Psi (r_1,r_2,\ldots r_k)$, we call its symmetrization a function of the form $a\sum\limits_{\pi}\Psi (r_{\pi (1)},r_{\pi (2)},\ldots ,r_{\pi (k)})(-1)^{\sigma (\pi )}$, where the summing is spread on all permutations $\pi$, $\sigma (\pi )$ denotes in the case of fermions the parity of permitation $\pi$, and 1 in the case of bosons. The stogare of a wave function in the form of tensor product  
\begin{equation}
|\Psi\rangle_{ind}=|\Psi_1(\bar r_1)\rangle\bigotimes |\Psi_2(\bar r_2)\rangle\bigotimes\ldots\bigotimes |\Psi_k(\bar r_k)\rangle
\label{non}
\end{equation}
is much more efficient than in the form $\sum\limits_{\bar i}\la_{i_j}|j\rangle$, because in the last case the summing is spread on the exponential number of summands. After the symmetrization (\ref{non}) we obtain the wave function in the form  
\begin{equation}
\frac{1}{\sqrt{k}}{\cal D}(|\Psi_1\rangle , |\Psi_2\rangle\ldots , |\Psi_k\rangle ; \bar r_1, \bar r_2, \ldots ,\bar r_k)
\label{sym},  
\end{equation}
where ${\cal D}$ is the determinant or the permanent (dependingly of the type of symmetry of the system - fermionic or bosonic) which is built on the wave functions and coordinates. 
We denote by $Sym|\Psi\rangle$ the result symmetrization of a wave function $|\Psi\rangle$ of fermionic or bosonic type. The function (\ref{sym}) can be represented as $Sym(|\Psi\rangle_{ind})$. This symmetrization can be applied to any wave function, in particular to those which are represented in the qubit form, where it means the computations of determinants or permanents of the amplitudes $\la^s_j$, where $s$- is the number of particle, $j$ - is the number of basic state. We consider here only fermionil ensembles. Since the computation of determinants for given values of coordinates has the polynomial complexity of te total number of particles, the presence of symmetrization in a simulation does not lead out from the framework of effective algorithms. 

The functions of the form $Sym(|\Psi_1(\bar r_1)\rangle\bigotimes |\Psi_2(\bar r_2)\rangle\bigotimes\ldots\bigotimes |\Psi_k(\bar r_k)\rangle )$ are called the functions of zero range of entanglement. These functions from the algebraic viewpoint are entangled because they cannot be represented as tensor products. But the storage of such functions does not require any substantial additional memory in comparizon to the non-entangled functions (\ref{non}), justifies these name. 

A representation of wave function of the form
\begin{equation}
|\Psi_{can}\rangle = Sym (\sum\limits_{j\in J}\la_j|\Psi_j\rangle )
\label{w}
\end{equation}
is called a canonical representation if the following conditions are satisfied:
\begin{itemize}
\item All the states $|\Psi_j\rangle$ are $k$- particle mutually orthogonal normed states. 
\item Each state $|\Psi_j\rangle$ has the representation of the form $\bigotimes_{h\in H(j)}|\Psi_{j,h}\rangle$, where for each $j\in J$ either $H(j)$ consists of one element only and $|\Psi_{j,h}\rangle$ is a basic state of our system (each particle in some point) of the form $|\bar r\rangle$, or $H(j)$ consists of $k$ elements $h_1(j),h_2(j),\ldots ,h_k(j)$ and each $|\Psi_{j,h}\rangle$ is one particle normed wave function of the form $\sum\limits_l\la^{j,h}_l|l\rangle$.
\end{itemize}
The algorithmic approach imposes the polynomial restriction to the number of elements in the set $J$. By virtue of the first condition of orthogonality for any spatial (and spin) configuration $|\bar r\rangle =|r_1,r_2,\ldots ,r_k\rangle$ of the considered system if $\langle \bar r\ |\ \Psi_{can}\rangle\neq 0$, then  there is no more than one value of $j$, such that 
\begin{equation}
\langle \bar r\ |\ \Psi_{can}\rangle =\la_j\ \la^{j,h_1(j)}_{r_1}\la^{j,h_2(j)}_{r_2}\ldots\la^{j,h_k(j)}_{r_k}
\label{v}
\end{equation}
We then can choose such combinations of values for $j$, and $r_1,r_2,\ldots ,r_k$ that the module of amplitude of a basic state $\bar r$ в $|\Psi_{can}\rangle$ $\langle \bar r\ |\ \Psi_{can}\rangle$ is not less than the chosen for computations value $g$ of the amplitude quantum. The choice of such a combination can be done in the logarithmic time of $1/g$ independently of $k$ if the number of elements in $J$ is fixed. Really, we have to search all directly written amplitudes in the state (\ref{w}) in descending order of their modules; the amplitide resulted from the multiplicationb in (\ref{v}) decreases exponentially and we reach $g$ in the logarithmic time. We thus can search in the polynomial time of $1/g$ all sufficiently large amplitudes in the canonical representation of state, despite of that the simple expansion of tensor product even with the fixed $g$ gives the representation of the length growing exponentially with the number of division points in the configuration space. 

Let we are given a canonical representation of space of a range $d$ of the form (\ref{w}). A canonical representation of a range $d+1$ for this state can be obtained as follows. We choose some spatial configuration $r^0_1,r^0_2,\ldots ,r^0_k$ along the rule defined above so that the corresponding basic state is not a simple summand in (\ref{w}). Then it corresponds to some value $j$. Let for each $s=1,\ldots k$ the function $|\Psi_{j,h_s(j)}\rangle$ have the form 
$ |\Psi '_{j,h_s(j)}\rangle+\la^{j,h_s(j)}_{r_s}|r^0_s\rangle + |\Psi ''_{j,h_s(j)}\rangle $ where $|\Psi '_{j,h_s(j)}\rangle$ (and $|\Psi ''_{j,h_s(j)}\rangle$) - are the summands which contain all the preceding (all the subsequent) to  $r_s^0$ values of coordinates of one particle. The representation of the function $|\Psi_{can}\rangle$ in the form of a state of range $d+1$ is obtained if we replace in (\ref{w}) the summand $|\Psi_j\rangle$ by the expression 
\begin{equation}
\begin{array}{ll}
\left(\sum\limits_{s=1}^k\right.&\left.\la^{j,h_1(j)}_{r_1^0}\la^{j,h_2(j)}_{r_2^0}\ldots\la^{j,h_{s-1}(j)}_{r_{s-1}^0}
|r^0_1,r^0_2,\ldots ,r^0_{s-1}\rangle\bigotimes (|\Psi'_{j,h_s(j)}\rangle +|\Psi''_{j,h_s(j)}\rangle )\bigotimes 
\bigotimes\limits_{b=s+1}^k|\Psi_{j,h_b(j)}\rangle\right)\\
&+\la^{j,h_1(j)}_{r_1^0}\la^{j,h_2(j)}_{r_2^0}\ldots\la^{j,h_{k}(j)}_{r_{k}^0}
|r^0_1,r^0_2,\ldots ,r^0_{k}\rangle .
\end{array}
\label{z}
\end{equation}
We thus consider the same wave function in the different forms. It follows from (\ref{z}) that this replacement gives the canonical representation of it. Such values of $j$ for which $H(j)$ consists of one element are called the main values. A main value of $j$ corresponds to certain values of coordinates of all particles $\bar r(j)$. We can fulfill the minimization of energy by the varying of amplitude corresponding to this value; this minimization results in the change of $\la_j$ and the corresponding renormalization of the rest amplitudes $\la_{j'}$ where $j'\neq j$; here all wave functions $|\Psi_{j',h_s(j')}\rangle$ remain unchanged. We start the process of energy minimization with the states of zero range. In each step number $d$ we have a state of range $d$, which energy is minimized by the alternate fixation of all main values of $j$ and varying of the corresponding amplitudes. After that we choose the different representation of this state of range $d+1$ for which the minimization of energy by the help of new main value relults in the change of state, etc. This process allows to minimize energy so that at each step we use the most economical representation of wave function. If a step $d$ does not already give the decreasing of energy in the passage to states of range $d+1$ we assume that an eigenstate is found for a given value of amplitude quantum. Let $	E_d^1,E_d^2,\ldots ,E_d^{f_d}$ be all energies obtained by the sequential minimization up to a range $d$ starting from a state of local minimun of energy $E_{d'}$ in a range $d'<d$, we call the values $E_{d'}-E_d^f$ energy defect. The values of energy defects characterize the influence of complexity of entanglement to the energy of the corresponding states for a given type of interactions. 

\subsection{Direct simulation in the form of secondary quantization}

The direct application of the method described above is difficult due the huge dimensionality of Hamiltonians in the coordinate form. This scheme is much easier to implement for the wave functions represented in the form of secondary quantization. We say that a particle with wave function $\Psi_j$ belongs to $j$-th energy level. In this case a function $Sym(|\Psi_1(\bar r_1)\rangle\bigotimes |\Psi_2(\bar r_2)\rangle\bigotimes\ldots\bigotimes |\Psi_k(\bar r_k)\rangle$ is denoted by $|\bar n\rangle =|n_1,n_2,\ldots ,n_L\rangle$, where $n_j$ equals to the quantity of such $l$, for which $i_l=j$ (population of $j$ energy level). Such functions form the orthonormal basis of space of states. The general form of wave function will be
\begin{equation}
\sum\limits_{\bar n}\la_{\bar n}|\bar n\rangle .
\label{sec}
\end{equation}
A Hamiltonian in this space has the form
\begin{equation}
H=\sum\limits_{k,l}v_{k,l}c_k^+c_l+\frac{1}{2}\sum\limits_{k,l,m,n}v_{k,l,m,n}c_l^+c_k^+c_mc_n
\label{hamsec}
\end{equation}
where operators of creation and annihilation of a particle in an an energy level $j$ have the form
\begin{equation}
\begin{array}{lll}
&c_j^+|n_1,n_2,\ldots ,n_j,\ldots \rangle &=(-1)^{\sigma_j(\bar n)}(1-n_j)|n_1,n_2,\ldots ,n_j+1,\ldots \rangle ,\\
&c_j|n_1,n_2,\ldots ,n_j,\ldots \rangle &=(-1)^{\sigma_j(\bar n)}n_j|n_1,n_2,\ldots ,n_j-1,\ldots \rangle ,\\
&\ \ \ \ &\sigma_j(\bar n)=n_1+n_2+\ldots +n_{j-1},
\end{array}
\end{equation}
for fermions, and $\sigma_j(\bar n)=1$ for bosons, and matrix elements have the form
$$
\begin{array}{lll}
&v_{k,l}&=\langle \Psi_k\ |\ \frac{p^2}{2m}+V_1\ |\ \Psi_l\rangle\\
&v_{k,l,m,n}&=\langle \Psi_l, \Psi_k\ |\ V_2\ |\ \Psi_m,\Psi_n\rangle
\end{array}
$$
and can be found by the integration on the spatial degrees of freedom and summing on spins by the standart rules (here the conjugation of tensor product changes the order of its components, $V_1,\ V_2$ are one and two particle potential, $p$ is the impulse operator. 
  
In this notations our process of energy minimization looks as follows. We start with some state $|\bar n\rangle$, which depends on a choice of functions $\Psi_j$, which must be orthonormal. By small variations of eigenstates $\Psi_j$ we reach a local minimum of energy of a state $|\bar n\rangle$ for some choice of these functions $\Psi^0_1,\Psi^0_2,\ldots ,\Psi^0_L$. After that we fix this basis in the space of occupation numbers and begin the further minimization of energy passing to nontrivial linear combinations of the form (\ref{sec}). At each step $d$ we have a function of the form $|\Psi_d\rangle=\sum\limits_{\bar n}\la_{\bar n}^d|\bar n\rangle$. For every $\bar n$ in the order of decreasing of their amplitude modules we fulfil the minimization of energy along all directions of the form $|\bar n\rangle+\la |\bar n'\rangle$, for which $|\bar n'\rangle=c_k^+c_l^+c_mc_n|\bar n\rangle$ for some combinations $k,l,m,n$. The resulting state will be $|\Psi_{d+1}\rangle$. The iteration of such steps gives a local minimum of energy, and correspondingly, an eigenvector of the Hamiltonian (\ref{hamsec}). A value of decreasing of energy in comparizon with the basic state is an energy defect.

We show roughly what algorithm for a hydrogen molecule can be obtained from this approach. This is the system consisting of two protons and two electrons that move in the alternating external electromagnetic field. The detailed consideration of this problem requires the account of the field dynamics that obeys Maxwell equations. We will not consider it in full generality and neglect the deposit of spins and vector electromagnetic potential into energy. We thus consider the Coulomb interaction only and spins are involved only through Pauli principle. The distribution of particles to the levels depends on the configuration of a system and it can change in time accordingly to our rules. For example, in the reaction of joining of two hydrogen atoms to the molecule initially the third level particles are electrons and protons, the second level particles are the pairs proton + electron, the first level particle is the object consisting of these two atoms e.g., the future molecule. Since protons are much more massive than electrons we can assume that at the begining protons are the particles of the second level, and electrons are the particles of the third level where each of them belongs to the tier of its proton. In the stationary state of hydrogen molecule the hierarchy looks otherwise. Electrons will be the third level particles, the second level particles will be: each of the protons and the pair of electrons, the first level particle - the molecule itself. If we neglect the photon emission the simulation of such a system looks as follows. 
For a given arbitrary but fixed position of protons we fulfil one step of finite-difference method of the electron dynamics simulation. We fulfil this step for each position of protons where the division points are distributed accordingly to the rule formulated above. We then do one step of the proton dynamics simulation by finite-difference method. This two step procedure is then iterated. Since the proton part of the wave function will change much slower than the electron part, we can fulfil many steps of electron simulation for a fixed proton position. Here the herarchy can either remain unchanged, or change - dependingly of the initial conditions. In the first case the simulation gives the endless complex oscillations of four particles.  

Now we will not neglect the emission of photon, e.g., consider the problem in the more generality. For a separated hydrogen atom we assume that its proton is fixed and comsider the electron dynamics. For basic states $A_j$ we take the energy electron eigenstates and the space-time photon states. For example, the process of emission of a photon by the electron that is initially in state $2s$ will look as the sequence of states of the form

\begin{equation}
S_1, \ S_2, \ldots , S_j, \ldots ,
\end{equation}

where each joint state $S_j$ of the atom and photon has the form

\begin{equation}
\begin{array}{ll}
&S_j=\frac{1}{\sqrt{j}}(|\Psi_{2s}\rangle\bigotimes |\phi_0\rangle +\sum\limits_{r=1}^{j}|\Psi_{1s}\rangle\bigotimes |\psi_r\rangle ,\\
&|\psi_r\rangle  =\exp (i\phi_r)\Omega (c\ r\Delta t)\bigotimes (|0\rangle +|1\rangle )
\end{array}
\label{emission}
\end{equation}
where $|\phi_0\rangle$ is the vacuum state, $\phi_r$ is the phase factor, $\Omega (R)$ is the characteristic function of spherical layer of the radius $R$, $c$ is the speed of light, and the last factor corresponds to the polarization. The photon energy is thus exactly determined and the time of emission is completely non-determined. Nevertheless, it follows from the representation (\ref{emission}) that the probability of emission converges to 1 if the time goes to infinity. Indeed, by virtue of the wave function reduction we have to choose one summands from 
 (\ref{emission}) with the equal probability. One step of electron movement simulation corresponds to the numerous steps of the photon simulation because just the photons create the potential determining the charged particle dynamics. 
That is even for the small number of iteration of the finite-difference scheme for electrons $j$ will be sufficiently large for that we can assume that the emission has been happened and the aton is in state $1s$. The system of two electrons in the field of two protons is considered analogously, so that we conclude that always the ground state of electrons must be considered if only there is no external field and the movement of protons is negligible. 

Considering the joining of two hydrogen atoms to the molecule we thus assume that the both electrons are initially in state $1s$. The change of hierarchy looks roughly as follows. When the protons close in the electrons loose the rigid constraint with their protons and lift in the hierarchy to the level 2. Their common state becomes strongly entangled due to the symmetrization and their pair can be considered as the new particle of the level 2, where electrons themselves becomes the particles of the level 3, that completes the forming of new hierarchy of the stationary hydrogen molecule state.  We note that this hierarchy is very convenient for the finding of electron pair eigenstates: the initial point of the coordinate system is placed to the middle of segment connecting protons. The time inversion gives the inverse process: the dissociation of hydrogen molecule resulting from the photon adsorbtion. Our approach embraces all known types of movements in the hydrogen molecule including its forming and decomposition, oscillations and rotations. The forms of spectrum can be thus found that corresponds to all these types of movements. But our method also embraces the movements of molecule that cannot be described in terms of classical dynamics. These movements result from the entenglement between all these four particles. Algorithmic approach is thus more general than analytic. 

The description of quasi-particles represents the special problem that arises in the simulation of systems similar to a crystal, that consist of a big nimber of particles. We do not touch this problem here. 

\subsection{Effects following from the algorithmic procedure of reduction}

The algorithmic reduction procedure of a wave function (in what follows - AR) is a nulling of the too small amplitudes. This procedure is principally different from the conventional collaps of a wave function in that AR gives the classical urn scheme and Born law of quantum probability, whereas the collaps does not; thus the collaps is included to the quantum theory as an axiom. Born law can be treated as the main "effect" following from AR. This "effect" is not the single. The AR procedure gives immediately the classical description of a dynamics if the corresponding Lagranjian action along the considered trajectories is large in comparizon with Planck constant (it has been mentioned by Feynman in (\cite{FH}) without explicit using od AR procedure, see also Appendix 1 of this work). AR thus gives the automatic passage from quantum to classical dynamics so that it is not necessary to take care of it when programming. If we consider a particle in two close potential holes with the high barrier then in the algorithmic approach no tunneling happens because it is blocked by AR, whereas in the conventional theory tunneling takes place for any barrier. This interesting effect of "blocking" of the quantum properties must be amplified in the many particle case in the passage to entangled states which we treat as the subordination of particles to the same particle of a smaller level. Really, if two particles are not entangled then each of them are described by its state vector of the form
$|\Psi_k\rangle =\sum\limits_{j=0}^N\mu_j^k|\phi_j^k\rangle ,\ \ k=1,2$. If they are entangled, its state is common and has the form $|\Psi_{com}\rangle=\sum\limits_{j,j'=0}^N\la_{j,j'}|\phi_j^1\rangle\bigotimes |\phi_{j'}^2\rangle$. Here for each fixed of the second particle state $j'$ the vector of state for the first particle will be defined with the resolution $N$ times less than for independent particles. This decreasing of the resolution represents the "blocking" of quantum states in entanglement. For example, it can lead to the impossibility of tunneling that is peculiar to independent particles, and as a corollary - to the stability of the many particle states that are defined by the classical method - as a minimal potential energy but not as a groundstate of the Hamiltonian. E.g., the situations are possible when some parts of complex quantum strongly entangled system can be better described by the classical means. 

\subsection{Some remarks}

We proceed with the general comments on the practical realization of the method on the real computers that impose the more severe limitations on the computational recourses than the abstract algorithms considered above. 
The methods of finite differences are applied here to one particle only. In the computation of one step of the unitary evolution for a particle of level $m$ we assume that all particles of the major levels are fixed in the space, whereas the spatial positions of the minor particles subordinate by hierarchy are averaged by the quantum law. In the molecular simulation we can treat the nuclei of atoms, electrons and photons as particles of the zero level, atoms as particles of the first level and molecules and ions as particles of the second and the next levels. For the most cases we can limit our consideration by the first three levels of hierarchy. As mentioned above, the states of electrons consisting in the same tier must be symmetrized by the Fock-Slater method. One more assumption about the almost unitary segments of evolution we can assume to simplify computations. In many cases when the accuracy of the photon wave function description can be neglected we can assume that electrons in the time of unitary evolution are either in the states with certain energy or instantly move from one of such states to another and emit or absorb the photon accordingly to the law of conservation of momentum\footnote{Of course, when computing the electron impulse the impulses of all enveloping particles must be taken into account.}. The electrons thus can travel between the energy levels permanently only in the lifting or lowering in the hierarchy or if they are considered as not entangled particles. This assumption means that we neglect the form of photon wave functions. In all likelihood\footnote{Though it is not exact fact.} such a consideration is sufficient for the majority of molecular processes, even for those which are substantially connected with the emitting and absorption of single photons. 

It is obvious that the computations cannot be performed in the real time mode of showing the "film". Hence even for isolated problems we should use the databases for electron states for fixed nuclei, the databases for photon radiation rate for all possible electron energy levels and the databases for the problems of simple dispersions (no more than 3 particles). These databases can be dynamical e.g., they could be formed in course of one process simulation and then discarded. But there are some databases which must be stored and gradually specified. For example, the databases for the stationary positions of nuclei in molecules and crystals and the corresponding states of electrons which form valence bonds and Brillion zones, and nuclei in the superpositions of spatial states (protons in hydrogen bounds), intensity of emanation and absorption of photons of the different impulses and polarizations for electrons and nuclei transition between such states. Such databases can account only a small number of close charged particles. 

In addition, the description of the majority of movements will be factually classical (especially it is true for the massive particles as nuclei); hence electron states must be often treated as classical potentials which determines the classical interactions of nuclei only, as in the Born-Oppenheimer approximation. It makes sense then to compose the databases of classical potentials created by such electron states (and, may be, the states of tunneling nuclei). For the nuclei consisting in the stationary molecules these classical potentials have the form of elastic potential 
 $kx^2$, where $x$ is the spatial coordinate $k$ is the constant determined by the model of local structure; thus only such constants $k$ must be stored. 

\section{Conclusion}

We outlined the general ideas of the algorithmic approach to physics that is based on the fundamental notion of an effective classical algorithm. The key assumption of this approach is the possibility to simulate a system of arbitrary complexity on classical computers with the polynomial computational burden. This approach does not contradict to anything established in experiments up to nowadays, but forbids the existence of a scalable quantum computer, which is allowed in the conventional quantum physics. We have described the approximate form of a classical algorithm designed for the simulation of systems for which quantum effects play an important role. This algorithm is based on the hierarchical representation of quantum states for a many particle system, when the whole tier of particles of the same level is in the entangled state and is treated as one particle of the next level. A unitary evolution of such a system is simulated by one particle quantum dynamics only. Transitions of the individual particles between the levels in this hierarchy are admissible and it makes possible to simulate chemical reactions. The efficiency of this algorithm is guaranteed by the procedure of reduction that eliminates all the states in superpositions which module of amplitude is less than some fixed value called an amplitude quantum. 

The rules we have formulated for the simulation of quantum dynamics exactly express the conventional quantum mechanical description through tensor products of Hilbert spaces with only one restriction: we null all too small amplitudes. In the framework of this limitation we account all effects resulting from quantum entanglement between all the particles in the considered system independently on their tiers. The division of the particles into tiers is only needed to economize the computational resources in the simulation by the possible using of peculiar one particle tricks.  

We saw that the reduction procedure that is treated as the nulling of too small amplitudes is in principle sufficient to the simulation of decoherence. This way makes possible to account all elenemtary events which probability is not less than $\frac{1}{T}$, where $T$ is the amount of the time we have at our disposal. This approach to the decoherence is very easy for programming and does not require any special description of the environment besides the evident fact that the computational resources can be distributed among the different parts of the physical space. The reduction immediately gives the classical urn interpretation of the quantum probability and Born rule for it that will be shown also in Appendix 1. At least the reduction transfers the quantum description of evolution to the classical without any additional suppositions that will be demonstrated in Appendix 1. 

The algorithmic approach thus gives the uniform description of a quantum dynamics without its division to the unitary dynamics and measurements; this description is also independent of the presence of an observer.

The further analysis of an amplitude quantum is not necessary for the construction of the simulating algorithm. Nevertheless we give the more vivid interpretation of amplitude quanta through the Feynman path integrals in Appendix 1. The idea of one possible way of representation of pseudo-Euclidean metric in space-time is described in Appendix 2. We underline that our approach in not an interpretation of quantum theory. It is rather the introductory part for the instruction on its practical implementation to the complex systems. Just as the particular tricks described in the both Appendices cannot be treated as the description of some mechanisms; it is the computational tricks only that are not the single possible. They touch two principles: the probability interpretation of wave function and the conservation of the pseudo-Euclidean metric in the transition from one inertial frame to another. These two principles cannot be reduced to more elementary things but they are known long ago in physics. It is shown how they can be represented in the framework of algorithmic approach if we do not introduce it to the model beforehand.

\nnn
{\LARGE \ \ \ \ \ \ \ \ \ \ \ \ \ \ \ Appendix\ 1 \\
Amplitude quanta in Feynman path integrals}
\nnn

The scheme of possible model for quantum evolutions shown above is based on the procedure of reduction which must be applied to each quantum state. Just this procedure guarantees the limitation of quantity of the summands in the notations of quantum states, and hence the effectiveness of this algorithm. This procedure can be assumed unconditionally in the algorithmic approach and just it factually imposes a ban on the creation of a scalable quantum computer, which is allowed in the conventional quantum physics. Nevertheless we show that this procedure can be made more sensible. Really, we have to limit somehow the minimal size of amplitudes of states in quantum superpositions, e.g., to introduce an amplitude quantum. We show the way which looks the most physical and which is based on Feynman path integrals. In addition, the description of quantum evolutions based on amplitude quanta is close to the classical and the transfer from the classical description to the quantum by this way looks very easy, whereas the invesre transfer is based on the reduction procedure only. 

Amplitude quanta were introduced in the work \cite{Oz2} on purpose to give a direct interpretation of Born's quantum formula for the probability in terms of a classical urn scheme, and this aim was reached. In this work we modify the notion of amplitude quanta in order to obtain the better simulation of quantum evolutions than the method of finite differences. Particularly, we require the easy transfer from the classical description of dynamics to the quantum and vice versa that is important for example, for the problems of molecular dynamics. The following condition is that the description of quantum dynamics must be independent of an observer, and the decoherence (that is the permanent soft measurent of quantum states) must be in-built in the model. Feynman path integrals (see. (\cite{FH}) is the form of quantum formalism which is the most appropriate for this aim. In this formalism the amplitude of passage of a particle from the point $1$ to the point $2$ is represented as the integral
\begin{equation}
K(2;1)=\int \exp\left(\frac{i}{h}S[x]\right){\cal D}x,\ \ S=\int\limits_{t_0}^{t^1}L(x'_t,x,t)\ dt
\label{Ker}
\end{equation}
over all possible trajectories $x(t)$, that go from $1=(t_1,x_1)$ to $2=(t_2,x_2)$, where the Lagranjian $L=E_{kin}-E_{pot}$ is the difference between the kinetic and potential energies; for example in case of a particle in scalar potential $E_{kin}=\frac{p^2}{2m}$ for impulse $p$, $E_{pot}=V(x)$. The function $K$ is called a kernel, or Green function (for the wave equation) and $S$ is the ordinary classical action along the trajectory $x$. 

Path integrals are convenient for us because they make possible to pass to the classical description of dynamics. The classical equation for trajectories has the form $\frac{\d S}{\d x}=0$; e.g., the small variations of trajectory do not change the action. It gives the simple practical rule for the passage from the classical description to the quantum and vice versa. Let we use a method of finite differences with the step $\Delta t$ for the solution of classical equations. We consider the element of action $\Delta S=L\Delta t$, corresponding to this step. If $\Delta S\gg h$, then the classical description gives the right picture; if $\Delta S\approx h$, we must pass to the quantum description. The initial distribution of coordinates can be taken as gaussian, so that the wave function in the initial instant has the form $\Omega (\bar x)\exp (i\frac{\bar p}{h}\bar x)$. And vice versa, in the quantum simulation the passage to the classical simulation must be fulfilled if $\Delta S$ becomes more than Planck constant $h$, because trajectories far from the classical give the deposits to the kernel that destructively interfere with each other\footnote{The element of action depends on the value $\Delta t$, which is not arbitrary. It cannot be very big because we then would not have the right method of finite differences. It imposes the restrictions to the area of applicability of the classical mechanics. But $\Delta t$ cannot be made arbitrary small as well because we then risk not to finish the quantum simulation at all.}. By virtue of our agreement to fulfil the reduction over all the states it leads to that when the action is large enough all the paths with nonzero deposit become classical. If we obtain the kernel by the formula (\ref{Ker}) then the wave function of our particle in a moment $t_2$ is expressed through this function in a moment $t_1$ by the formula 
\begin{equation} 
\Psi(t_2,x_2)=\int K(t_2,x_2; \ t_1,x_1)\Psi(t_1,x_1) dx_1.
\label{wavefunction}
\end{equation}
The particular form of Lagranjian is not important for us, the more complicated expressions are allowed, for example, it can depend on the second derivative of $x$ by time, - it requires only little extension of the internal memory of amplitude quanta. In the framework of Hilbert formalism for many particles the formulae (\ref{Ker}) and (\ref{wavefunction}) are true in the many particle case as well, provided by a trajectory we mean a trajectory of the correspondent many particle system. Now we turn back to the one particle case. We admit that the total number of such trajectories is limited from the very beginning and the corresponding particle is moving along each of these trajectories. These ficticious particles are called amplitude quanta (a.q.)\footnote{This definition makes possible to simulate the dynamics numerically but yet does not give the urn model of probabilities; to obtain such a model we must further split these a.q. to the more elementary a.q. (see below). }.
We could apply the collision model for a.q. to guarantee the chaotic character of a.q. movement. Changing of the regime of collision we could try to economize computational recourses required in the simulation. But in fact we need only to change somehow tarjectories from point to point chaotically, and for this the collisions are not needed. 
\nnn

{\Large{Bounded amplitude quanta}}
\nnn

We now consider a.q. in more details. An amplitude quantum is a point object which moves in three dimensional space. In each time instant (we assume that the time runs eaually in all the space) a.q. $\a$ has the dynamical parameters: coordinates, speed, denoted by $x(\a ),\ v(\a )$, and special parameters as which we in this section consider an amplitude denoted by $\la(\a)$, (phase and mass can be used instead of amplitude), and we call such a.q. bounded. In the next section we will consider free a.q. whose special parameter is a type which takes 4 values. 

A.q. are denoted by small Greek letters $\a, \b, \g$, the dynamical parameters of a.q. $\a$ are designed by $x(\a), \ v(\a)$, and the special parameters by $\tau(\a)$. At first we consider the simplest version of a.q. where $\tau(\a)=\la(\a)$ is the amplitude associated with $\a$. We assume that set of values for coordinates $x$ of a.q. consists of the nodes of grid with step $\e$ that can in general depend on the cooredinate. Let the coordinates of a real particle be measured with accuracy $\d=r\e,\ r$ integer\footnote{It is not convenient to assume that always $\d=\e$ because a wave function would then have a very discontinuous form. }.
One position of a real particle then corresponds to $(\d /\e)^3$ positions of a.q. All these positions fill the cube 
$C^{\d,\e}_{l,n,m}$, which consists of points of the form $l\d +\e j,m\d+\e k,n+\e s$, where $j,k,s\in\{0,1,\ldots,r-1\}$. Hence, given the positions of all a.q. the amplitude of that the real particle is in a cube $C^{\d,\e}_{l,n,m}$ is 
\begin{equation}
\sum\limits_{\a:\ x(\a)\in C^{\d,\e}_{l,n,m}}\la(\a).
\label{psi1}
\end{equation}
We denote by ${\cal K}_\e$ the set of all a.q. in the considered area with the account of their coordinates and speeds determined with the accuracy $\e$, and their special parameters in time instant $t$, where the lowest index will be often omitted. 
 
If we fixe a value of $\d$, we can obtain trough this formula the corresponding amplitude distribution, e.g., the wave function which is denoted by $|\Psi_{{\cal K}_\e,\d}\rangle$. The area of space where the density of a.q. is not vanishing is called the a.q. bubble corresponding to the considered real particle. A.q. are thus treated as identical copies of the real particle; these copies differ only in their coordinates and speeds. 

The first step is to establish how a.q. speeds and amplitudes must be transformed in the collisions a.q. (or the collisions with nodes of spatial grid) for that fro all $t$ $|\Psi_{{\cal K}(t)_\e,\d}\rangle$ is a solution of Shoedinger equation within $C(M)t\d^3$ ($M$ is the total number of a.q.), or in other words, for that in case of steady Hamiltonian with this accuracy the following equation is satisfied
\begin{equation}
|\Psi_{{\cal K}(t)_\e,\d}\rangle\approx\exp\left(-\frac{i}{h}Ht\right),
\label{evol}
\end{equation}
and in case of time dependent Hamiltonian his equation is satisfied again but in sense of chronolochronological exponential. 

Transformations of a.q. parameters in a.q.collisions (or collisions with nodes of grid) are represented in the form of reactions 
$$
\bar v_1,\ \bar x_1, \la_1, \ \D t_1; \ \ \bar v_2,\ \bar x_2, \la_2  \ar\ \bar v'_1,\ \bar x'_1, \la'_1; \ \ \bar v'_2,\ \bar x'_2, \la'_2, 
$$
where $\D t_1,\ \D t_2$ is the time past from the previous collision of the first and second a.q.
In view of formula (\ref{psi1}) the main role in the detremining of a quantum state play the numbers $\la(\a)$. We determin the transformation of these numbers in collisions as follows
\begin{equation}
\la'_j=\la_j\cdot e^{\frac{i}{h}\D S_j},\ \ j=1,2,
\label{evol-amp}
\end{equation}
where 
\begin{equation}
\D S_j= L_j\D t_j,\ \ L_j=E_{kin}-E_{pot}
\label{action}
\end{equation}
is the Lagranjian of $j$-th a.q. computed in the point lying in the middle of the way from the previous collision. New 
speeds can be obtained from the condition that all the collisions are elastic. Here the different variants are possible, for example we could assume that if amplitudes 
$\la_j$ interfere constructively the collision is less elastic and more similar to adhesion, if the amplitudes interfere destructively the collision is close to elastic. These tricks can economize the computational resources. In fact the simplest way is to assume that the speeds of all a.q. after collisions are distribured randomly and uniformly independent of the previous speeds of the colliding a.q. We then sum the numbers $\la(\a)$ for all a.q. $\a$, containing in the cube $C^{\d,\e}_{l,n,m}$ in time instant $t$. The obtained value of the function $|\Psi_{{\cal K}_\e,\d}\rangle$ is the required approximation of the real wave function of considered particle provided the trajectories of all a.q. are distributed randomly and uniformly among final points for all time instants. 
Really, in this case the computation of kernel by formula (\ref{Ker}) can be approximately represented as the summing of numbers $\la(\a)$ over all a.q. $\a$ contained in the cube corresponding to the point 2 provided all initial a.q. were 
in the cube corresponding to the point 1 and the sum of all 
$\la(\a)$ for a.q. in the initial instant is equals 1. Our procedure of finding $|\Psi_{{\cal K}(t_2)_\e,\d}\rangle$ is then an approximation of the formula (\ref{wavefunction}), e.g. it gives the approximation of wave function in the moment $t_2$ provided $|\Psi_{{\cal K}(t_1)_\e,\d}\rangle$ is an approximation of it in the instant $t_1$.

The formula (\ref{psi1}) is then the discrete analog of Feynman path integral (\ref{Ker}). The accuracy of this approximation is the more the less numbers $|\la(\a)|$ are. If we assume that the amplitude is grained, e.g. that there exists the amplitude quantum $g$, then the maximal accuracy is reached when $|\la(\a)|=g$. 
This situation is considered in the next section.  

The numerical experiments show that in most cases the collision model gives no economy and even slow down the simulation. The reasons for the collision model for a.q. are as follows. We consider the system of particles moving chaotically so that the single potential is the potential of collisions with the other particles\footnote{Strictly speaking it is yet not a Brownian motion. This motion is obtained if we consider collisions of a.q. with less sized particles that makes the trajectories of a.q. absolutely chaotical; it is not significant for our purposes.}. 
If the mean track length of particles is large in comparizon with their size and its total number is large then such a system can be treated as a Markoff process, namely as a model of Brownian motion (see. (\cite{Hi}))\footnote{Bashelier tried to investigate this model (about 1900), but the equation for the density of particles in Brownian motion was obtained by Einstein (1905); further such processes were treated by Wiener (1923-38) and Levi (1937-40).}.
Let $u(x,t)$ be the density of particles in point $x$ in instant $t$, then the density satisfies the equation of heat conductivity\footnote{Really, let $\phi(y,t)$ be the fraction of partoicles which shifted in a time  $t$ from $x$ to $x+y$. We assume that it has the dispersion $Ct$, and symmetry. From the definition of $\phi$ we then have $u(x,t+t_1)=\int_Ru(x-y,t)\phi(t_1,y)dy$ and now it is sufficient to expand $u$ to the row of degrees of $t_1$ and apply the both assumptions to obtain the equation of heat conductivity.}.
\begin{equation}
u_t=\frac{1}{2}Cu_{xx},
\end{equation}
where $C$ is a constant, and its solution for $u(x,0)=\d(x-x_0)$ has the form of Gauss curve
\begin{equation}
u=\frac{1}{\sqrt{2\pi Ct}}\exp\left(-\frac{x-x_0}{2Ct}\right) .
\label{brownian}
\end{equation}
On the other hand, let us consider the quantum evolution for a free one dimensional particle which is at rest initially and has the Gauss amplitude distribution proportional to $\exp\left(-\frac{\a}{2}x^2\right)$. If we find its kernel by the formula (\ref{Ker}) then we obtain (see (\cite{FH}) for the details):
\begin{equation}
K(x_2,t_2,x_1,t_1)=\left(\frac{2\pi \ ih(t_2-t_1)}{m}\right)^{-1/2}\exp\left(\frac{im(x_2-x_1)^2}{2h(t_2-t_1)}\right).
\label{Kerfree}
\end{equation}
substituting it to the formula (\ref{wavefunction}) and calculating the wave function of free particle we can find the density of probability to find this particle in a point $x$ in an instant $t$ it will be proportional to 
\begin{equation}
\exp\left(-\a x^2\frac{1}{1+\frac{\a^2h^2t^2}{m^2}}\right).
\label{wavefunctionfree}
\end{equation}
Comparing formulas (\ref{wavefunctionfree}) and (\ref{brownian}) we conclude that all densities have Gauss form but the density of Brownian particles spreads faster in small times than the bubble of a.q. for free particle. 
Really, if the both densities are equal initially then for small $t$ the coefficient of the exponent for Brownian particles has the form $1-At$ whereas for a.q. it has the form $1-Bt^2$ with positive $A$ and $B$. It would seem that it testifies against Brownian model for trajectories of a.q. 
But this drawback can be corrected.  The point is that in course of reactions between a.q. the permanent redistribution of a.q. will go between the different areas of simulated space. If an amplitude quantum did not collide for a long time we agree to redistribute its amplitude among the other a.q. proportionally their own amplitudes\footnote{We can eliminate such a quantum and create the new one with the same speed and coordinates corresponding to the amplitude distribution.} (the analogous action takes place in case of free a.q.).
It makes possible to prevent too fast spread of a.q. bubble. The similar procedure is necessary for the description of electromagnetic field where photons are emitted by the charged particles. Imagine that the a.q. bubble has the shape of sphere which radius varies in time: $\{ r:\  |r|<R(t)\}$, where the redistribution and reactions go in the spherical area on the border of the sphere $\{ r:\  R(t)-\e <|r|<R(t)\}$, where $\e>0$ is sufficiently small and no reactions between a.q. happen inside the area $\{ r:\  |r|<R(t)-\e\}$. What would happen with a.q. density in such a redistribution ? The density on the considered narrow border will correspond to the wave function of the particle. Since this density has Gauss form everywhere, it will be equal to 
$|\Phi(t)|^2$ inside the bubble despite of that the reactions and redistributions go on the border only. The collision dymanics thus guarantees the required a.q. density almost without the reactions and redistributions - they are needed on the border of the bubble only. The same reasoning are true for a free particle with the nonzero mean impulse. 
This qualitative reasonings show that the collision model is in principle good for Feynman trajectories at least for free particles, because this model maintains the right form of density itself and it could proposedly save the computational resources comparatively with the seach of trajectories by various methods like Monte-Karlo. 
Our reasonings are approximate because we account a.q. only quantitatively, whereas the wave function is obtained 
from a.q. through the formula (\ref{psi1}). But in case of free a.q. their quantitative density will better correspond to the wave function that makes our reasonings about the collisions more sensible. We note that in a.q. model we can assume that all modules of amplitude parts $\la_{\a}$ of a.q. are equal and they differs in phases only that makes interference picture after the summing. 
\nnn

{\Large{Simulation with a.q.}}
\nnn

A.q. model is based on the a.q. dynamics, their redistribution and the computation of the wave function. We describe these steps sequentially. 
 
At first we make one important remark. We have assumed to divide the model into two parts: the first is accessible to users and the second accessible to the administrator only. For example, we denote the user (physical) time by $t$, and the simulation time (the time of administrator, or computer time) by $\tau$. It can be understood as if a user observes the "film" that is shown to him by the administrator and in this "film" the time flies accordingly to the scale $t$. But the creation of this "film" requires the other time; and just this time which has sense for the administrator only and which is equal to the quantity of steps in the simulating algorithm we denote by $\tau$. The "film" is interactive and a user can interfere in it, the time scale $\tau$ will be then torn and the simulation must be started anew. The part of model accessible to a user must correspond to the real observable world and we shall consider it in this section. The main notion here is the space. The space must be represented as the three dimensional grid with the small step $\D x$ where all directions are equivalent; we thus admit only such spatial positions that coinside with the nodes of this grid. 
The movement of poinwise particle is then represented as sequential jumps from one node to the neighboring. We do not admit the skip over several nodes because it would make impossible to detect a particle in the intermediate positions \footnote{There is one more reason against such skips over nodes. If the speed is high and the trajectory is close to some singular point where the potential converges to infinity then in case of such skips the movement of particle would depend not on its initial position only, but on the initial time instant that is unacceptable.}. 
We then obtain that the time of simulation $\D \tau$ will be proportional to the distance overpassed by a particle in the physical time $\D t$. That is the same physical time frame $\D t$ requires the different costs of computer time $\D \tau$, that is proportional to the speed of particle\footnote{We do not consider relativistic effects here.}. 
Imagine that the space is potentially infinite: we are always able to assemble the new nodes to the grid. For the simulation of a particle moving with the speed $v$ in the fixed time frame $t$ it is thus required the simulation time $\tau$ proportional to $v$. It means that we cannot simulate arbitrary speeds. The physical speeds accessible to the simulation must be limited by the value proportional to the maximal admissible time of waiting of the "film" $\tau_{max}$ (and of course to the frequency of the simulating processor). 

\nnn

{\Large{Coordinate and impulse representation of wave function in terms of a.q.}}
\nnn

We consider the passage to impulse representation of wave function in therms of a.q. 
Let we are given a reservoir with a set ${\cal K}$ of a.q. such that the corresponding wave function of particle $|\Psi_{{\cal K}_\e,\d}\rangle$ is computed by the formula (\ref{psi1}). We suppose that 
\begin{itemize}
\item The value $\e$ is so small that there are a lot of small cubes of the size $\e$, that compose the connected massive and for which the values of wave function found by the formula (\ref{psi1}) are close. 
\item Due to the chaotic character of collisions in the chosen model we can assume that the a.q. impulses are distributed by the Gauss law inside of each small cube. 
\end{itemize}
We choose an arbitrary value of impulse of the particle: $k_0$, and show how to find the value of wave function of a given state $|\Psi_{{\cal K}_\e,\d}\rangle$ in the impulse representation. 
The conventional way is ti apply Fourier transform:
\begin{equation}
\Psi(k)\rangle=A\int\limits_R\Psi(r)e^{-irk}dr.
\label{Psi2}
\end{equation}
Since each a.q. has not only the coordinates but also the speed, e.g., impulse, we can introduce the new amplitudes $\mu$ by the phase shift of amplitude parts of the quanta:
\begin{equation}
\mu_{\a}=\la_{\a}e^{-ir_{\a}k_{\a}}.
\label{shift}
\end{equation} 
Substituting into (\ref{Psi}) the expression for the wave function from (\ref{psi1}) and using (\ref{shift}) we obtain
\begin{equation}
\Psi(k_0)\rangle=A\sum\limits_{\a}\mu_{\a}e^{ir_{\a}(k_{\a}-k_0)}.
\end{equation}
Due to our assumption about the distribution of a.q. impulses we obtain that in this sum all the summands corresponding to such $\a$ that $|k_{\a}-k_0|>\epsilon$ for sufficiently small $\epsilon$ give the negligible deposit and we can write the approximate equation
\begin{equation}
\Psi(k_0)\approx\sum\limits_{\a:\ |k_{\a}-k_0|<\epsilon}\mu_{\a}.
\label{imp}
\end{equation}
It means that it is easy to pass from the set ${\cal K}$ of a.q. to the coordinate representation of wave function by (\ref{Psi2}) as well as to its impulse representation by (\ref{imp}); in the both cases this is the simple summing of a.q. of the special form. The single action we need to fulfil to obtain the impulse representation is the phase shift (\ref{shift}) (for the symmetry we could include the half of this shift to the a.q. type the complexity of computation of the coordinate and impulse representations will be then the same. 

\nnn

{\Large{Amplitude quantum representation of many body systems}}
\nnn

The passage to the system of many particles in the amplitude quanta representation is natural. Let $S_1,\ldots, S_k$ be the particles in the same tier (usually $k=2$). We assume that some lists of the form 

\begin{equation}
\a_1 ,\ldots \a_k
\label{comp-qa}
\end{equation}
 a.q. of these particles form an object called amplitude quantum of the whole system. A trajectory for such an a.q. is then defined naturally; collisions are defined as the events when these complex objects occur in some area of the configuration space, etc. The remark about the density of the division points remain valid for the many particle case as well.

For such complex a.q. the rule of transformation of amplitude parts is defined again accordingly to (\ref{evol-amp}), where the action along a given path of such a quantum is defined as usual taking into account the kinetic energy of all its components and its potential energy. If the total number of a.q. is large this model gives the same dynamics as Shroedinger equation for many particles. 

Let us consider how Born rule can be derived from the a.q. approach. Suppose, for example, that we measure the spatial position of one electron. It means that the contact between this electron and a many particle system takes place when the complex a.q. of the form (\ref{comp-qa}) arise. For the distinctness let $a_1$ denote an electon a.q., and the other elements of the list denote a.q. of particles contained in the measing device (including photons) so that there such particles among them which point to the measured position of the electron - denote them by $a_2,\ldots, a_s, \ s<k$. As usual we assume that the modules of amplitude parts of all complex a.q. are approximately equal and are close to the value of amplitude quantum $\e$ that is essentially less than amplitudes of a separate electron. The own evolution of the electron can be then neglected because the measurement of its coordinates presumes very intensive interaction with the measuring device; and we must simulate the evolution of the complex system electron + measuring device. Let we be interested in the hit of electron to some volume $\Delta V$ with the coordinates of centrum $x$. Among all complex a.q. there are such for which the position of electron a.q. $a_1$ is in  $\Delta V$. Their total number $l$, of course, does not connected with any amplitude because $a_{s+1},\ldots ,a_k$ take arbitrary values. But due to the huge total number $l$ of complex a.q. and that their module of amplitudes are close to the minimal value, $l$  must be proportional to $|\Psi (x)|^2$, because the simulated evolution is unitary\footnote{It is reached after the numerous redistributions of a.q. in the many step simulation of quantum evolution.}. It means that if we choose arbitrarily a complex a.q. of the system electron + measuring device, the probability to obtain the position of measuring device corresponding to the target electron position will be proportional to $|\Psi (x)|^2$, with the factor independent of $x$. This is the classical urn scheme giving Born rule for the quantum probability. 

In the framework of a.q. method it is not easy to represent a measurement of many particle system accordingly to the Hilbert formalism picture, e.g. in the form of expansion of the unit operator to the sum of mutually orthogonal projections\footnote{We note that it is not easier to realize experimentally such an abstract measurement for many bodies.}.
It is easy to describe a measurement of one particle in an ensemble - it was done above. Such a measurement gives the information about the whole ensemble only if the particles strongly interact. For example, if we consider a measurement of position of atomic nucleus then the a.q. of nucleus disposed far from the first collision of a.q. of the same type in the measurement. It leads to the fast disappearance of a.q. of electrons disposed near disappeared nuclear a.q. that is we have the effect similar to the measurement of a state of the form  $|00\ldots\rangle+\la_1|11\ldots 1\rangle+\la_2|22\ldots 2\rangle+\ldots$. We can hope to get the description of EPR pairs which demonstrate the violation of Bell inequalities, namely - the fast change of the apparatus measuring one particle such that the light cannot get the other particle before the instant of measurement. For this we must use a.q. corresponding the different basises of measurement; but anyway, it is necessary to apply administrative signals which spread in the simulating system media and which cannot lead to the informational exchange between users. 

The interesting question: how to choose components of complex amplitude quanta of the form (\ref{comp-qa}) from one particle a.q. ? When a many body system is formed by touching of one particle bubbles, we can assume that such complex a.q. are formed in the sequentional collisions of one particle a.q., if the information about these collisions is somehow stored. To specify the regime of forming and dissociation of many particle a.q. we can use genetic algorithms. 

\nnn
{\Large{Permanent measurement as norming administrative signals}}
\nnn

The single not local procedure in the a.q. formalism is the annihilation and creation of a.q. that is introduced for the preserving of its total number. This procedure is similar to the norming of a wave function and it requires the signals which spread faster then a.q. can move\footnote{We could say: they spread in the other media which is accessible to the administrator but not to users.}; these signals are called norming. Norming signals can carry no information forethought by a user A to a user B. But these signals in the model are necessary for the explanation of quantum nonlocality established in the series of experiments\footnote{For example, see (\cite{As}).}. In terms of "films" the norming signals mean that such a film is prepared beforehand and its parts are demonstrated to a user in turn as they are ready. Here a user cannot make over already done parts but can order the following parts for a future. This preprogramming makes possible the simulation of quantum non-locality by a classical computational network which we called the administrative system. If only a user has rights to look inside it he would observe the impossible thing: signals travelling instantly. The absence of such rights of a user just means the limitation on the speed of information transfer. This limitation can be reformulated in terms of a "free will" as above. It has the fundamental nature and is connected with the prevention of logical paradoxes. 

One aim of the intorduction of a.q. is to find a first principle description of the current quantum state in course of its evolution that is impossible with the conventional Hilbert formalism. 
The chemical method of a.q. makes it possible. A bubble filled with a.q. is the basic model of one particle quantum states. To describe its dynamics we must know what is happening with the separate a.q. Our consideration here do not depend on what form of a.q. we use: free or bound. We have the source of a.q. - their collisions and it is needed only to describe the procedure of elimination of a.q. that will guarantee the stability of its total number. We agree to eliminate each a.q. which did not collide in course of sifficiently large time $t>t_0$ and to eliminate the colliding a.q. if they are mutually antithetic: $x^s$ and $x^{-s},\ x\in\{\a , \b\}$ ($r$-reduction).  This method is appropriate for one free partiucle moving in the space; but yet for a particle in a potential relief this method can lead to too fast decreasing of the a.q. total number due to their spreading on the large area. To prevent this undesurable process we will use a.q. recycling that is equivalent to the norming of wave function: the dissappearing a.q. will be redistributed to the other spatial positions accordingly to the amplitude distribution found by the formula 
(\ref{psi1}). An eliminated quantum must transform to its copy with the same impulse. For free a.q we could guarantee the determinicity by some simple trick that is not necessary. This procedure is similar to the norming of current state which is the subject of permanent soft measurements\footnote{If the dynamics of a.q. leads to the situation when there is no a.q. in some area then from the Hilbert spaces viewpoint it is equivalent  to the soft measurement of wave functin (see (\cite{Me}))}.

We have already mentioned that it is not necessary to introduce the special procedure of measurement to the algorithmic formalism. Newertheless, this trick can be not useless to speed up the preparation of the "film". Such a measurement happens in the moment of break-up of the bubble to the two disconnected parts\footnote{The  recognition of this moment can be based on the permanent transmission of the special value of connectivity from ane quanta to the other in its collisions. This value changes such that a component of connectivity is characterized by the same value of connectivity of all a.q. which belong to this component. 
In particular the connectivity means the same value of connectivity of all a.q. corresponding to this particle.}. 
In this case as a new bubble we take a component of connectivity in which the first collision of a.q. of the same type has happened; the rest a.q. are redistributed on the new bubble accordingly to the procedure described above\footnote{The slightly different method is possible when a part ${\cal B}_1$ of the bubble which has the larger surface loses a.q. faster because they fly away and these a.q. arise anew in the other area ${\cal B}_2$, that leads to the disappearance of ${\cal B}_1$; or a combination of such tricks.} . 

For the realization of permanent measurements it is needed to have norming signals with the instantaneous access to all a.q. of the considered particle. These signals have no physical sense for a user because it cannon carry any information put-up by a user. 
In contrast to a.q. that are the copies of one physical particle and which speed cannot exceed the speed of light, the signals spread instantly. It makes possible to represent not only the movement of particles in the field but also the behavior of the field itself, for example the experiments on the detection of EPR pair (see (\cite{As}), that could not be visualized without administrative signals. 
These signals are internal processes of the simulating system and they cannot carry any user's information thus it is compatible with the fundamental relativistic limitation to the speed f information transmission. 
It can be explained otherwise. What is called a "free will" exists among the users only, not in the world of a.q. and system signals, because in that world all including the results of observations are determined. Hence the signals which determin the shape of a.q. bubble can travel with arbitrary speed without violation of the relativistic ban on the superluminal transfer of information - there is no information without a "free will". 
Users can get an information about a.q. only through the measurements connected with a.q. collisions as was described above. This method of informational exchange between users of our imaginary system is authorized and it does not allow to transmit an information faster than a.q. move\footnote{But if we imagine that a user has somehow learnt the positions and speeds of all a.q. in the whole space and he has an instantly working computer, then he could transmit his messages with arbitrary speed.}. 
This approach can be applied for photons as well if we take into account the features of electromagnetic field. We note that the description of the relativism itself in terms of a.q. represents the separate task which lies beyond the framework of this paper.
\nnn

Summing up, we note that the simulating process for many body system consists of two types of a.q. transformations
\begin{itemize}
\item local reactions of chemical type, and 
\item nonlocal "norming" signals.
\end{itemize}
\nnn

{\Large{Free amplitude quanta}}
\nnn

The simulation with bound a.q. is based on the algebraic operations over binary notations of amplitudes that cannot give a classical urn model for a quantum probability. Here we show how such a model can be obtained if we split bound a.q. to ithe small summands called free a.q. 
In terms of free a.q. we can give the classical interpretation of a quantum probability without usage of algebraic operations and basing on the reactions of chemical type between a.q. only. Free a.q. express the grain of amplitudes whereas bound a.q. express completely the grain of space only. Using free a.q. we hope to obtain such effects as colective excitations in the simulation of many body systems, that cannot be obtained if we consider amplitudes as continuous. Free a.q. more correspond to the ideology of analogous simulation, not of the digital one. This is why bound a.q. are more convenient for practical simulation. Free a.q. makes possible to reduce all description of quantum dynamics to the reactions of chemical type; this is why we devote one section to free a.q.

Given an amplitude quantum $q$ we denote its type by $\tau (q)$. Each type has the form 
\begin{equation}
x^s_r,
\label{quantum}
\end{equation}
where $x\in \{\a,\b\}$ determins which part of amplitude is represented by this quantum: real ($\a$) or imaginary ($\b$),  $s\in\{ +,-\}$ determins the sign of this quantum and $r$ is the list of the form $r=j\ r_1\ \ldots \ r_k$. Here the first element $j=0,1,\ldots,N-1$ determins the basic state $|\Psi_j\rangle$ which this amplitude corresponds to and it varies accordingly to the coordinate of this quantum (see below) and the rest elements contain the auxiliary options of the quantum. 
We assume the conventional rules of handling with signs. 
We denote by $[ x^s_j]_{\cal B}$ the total number of a.q. of the form (\ref{quantum}) in the bubble ${\cal B}$, where in the lower index the auxiliary options and ${\cal B}$ will be often omitted. We put $ [x_j] = [x^+_j]- [x^-_j]$.

The result of all possible annihilations of a.q. of the types $x_r,\ x_r^{-1}$ is called $r$- reduction. 
We define real nonnegative numbers
\begin{equation}
p_j=\frac{ [\a_j]^2+[\b_j]^2}{ \sum\limits_{x\in A,\\ 0\leq k\leq N-1}[x_k]^2}.
\end{equation}
Such a number $p_j$ can be considered as a probability to obtain a real state of the form $(x^s_j,x^s_j)$ for some $x\in A,\ s\in\{ +,-\}$ as a result of all sequential $j$- reductions in the bubble ($j=0,1,\ldots ,N-1$), if a real state is treated as a result of collision of a.q. of the same type\footnote{Of course, this interpretation of Born rule is much worse than that we have done above, because it requires the "coupling" of a.q. of the same type that is an artificial construction; newertheless it fully answer to the spirit of chemical type reactions between a.q.}.

For a given state  
\begin{equation}
|\Psi\rangle=\sum\limits_{j=0}^{N-1}\la_j|\Psi_j\rangle
\label{Psi}
\end{equation}
of the considered system we denote ${\rm Re}\ \la_j,\ \Im\ \la_j$ by $\a_{j,\Psi},\ \b_{j,\Psi}$ where the lower index $\Psi$ will be often omitted. We represent a normed state (\ref{Psi}) by some bubble ${\cal B}$. 

The state (\ref{Psi}) is called the corresponding to a bubble ${\cal B}$ 
if and only if for all $j,k=0,1,\ldots,N-1$ and $x,y\in A$ the following equations take place
$$
\frac{[x_j]_{\cal B}}{[y_k]_{\cal B}}=\frac{x_{j,\Psi}}{y_{k,\Psi}}.
$$
In this case we write $|\Psi\rangle=|\Psi\rangle_{\cal B}$. Applying the conventional rules for calculation of probabilities we conclude that if $\Psi=\Psi_{\cal B}$, then for all $j=0,1,\ldots,N-1$, $\ p_j=|\la_j|^2$ that substanciates the probability interpretation of an amplitude squared module. 

The phase shift of the form $\Psi\ar e^{i\phi}\Psi$ is represented by the list of reactions between the colliding a.q. of the form\footnote{The type of the second quantum in collisions does not play any role and it is thus omitted. A collision is happened if the coordinates of the second quantum belongs to some volume around the first quantum, for which the reactions are written.}
\begin{equation}
\begin{array}{ll}
\b^s&\ar\b^s,\a^{-s},\\
\a^s&\ar\a^s,\b^{s},
\end{array}
\label{phase-shift}
\end{equation}
where $s\in\{ +,-\}$, and $\phi$ expresses the reaction rate and it dpends on a.q. density and their volume. 
To make $\phi$ independent from a.q. density we can vary the volume $v(\a )$ of each a.q. $\a$, so that $v(\a )=v^0_\a\d t$, where $d t$ is the segment of time passed from the previous collision of this a.q. with the a.q. corresponding to the same particle; $\phi$ will thus be determined by the value of $v_0$ only. We can allow for the valume the negative values as well that can be stored with their sign in the computer memory. We can allow also the negative signs for a value, that can be stored with its module in the computer memory. Here the amplitude of amplitude quantum with negative volume is obtained as usual but is taken with the sign minus (it is equivalent to that in (\ref{phase-shift}) we take $-s$ instead of $s$ in the right side of reactions). 
In case of free a.q. we then have to choose $v^0_\a$ for a quantum $\a$ such that this number is proportional to the element $\d S$ of the action for the quantum $\a$, that is calculated accordingly to the formula (\ref{action}). All the quantum evolution is thus simulated by the reactions of the form (\ref{phase-shift})\footnote{In order to make the model more symmetrical and not to separate explicitly the real and the imaginary parts of amplitudes we can intorduce these a.q. of the types $\a$ and $\b$ in the different basices of the algebra of complex numbers of the form $e^{i\phi_j},e^{i(\phi+\frac{\pi}{2})}$ for $\phi_j=2j\pi /N, \ j=0,1,\ldots,N-1$.}. The a.q. number in the reactions always grows that is compensated by their decreasing in the big distances because we agree to eliminate each a.q. which has no collisions with the others for a sufficiently large time.

The different approach is that we consider a.q. as pointwise objects and their collisions happen when they occur in the same small segments of the configuration space simultaneously. If the size of these segments goes to zero and the total number of a.q. - to infinity we obtain the wave function dynamics determined by Shroedinger equation. If we solve it by the finite difference method then the change of division points density accordingly to the rule $\rho (x)=C\ |\Psi (x)|^2$ expresses the most efficient expense of computational resources for this method. 

We see that free a.q. not only reduce the quantum probability to the classical urn scheme but also reduce the control over evolution for arbitrary complex Hamiltonian to the varying of a.q. sizes (just a.q. sizes depend on the potential), whereas the reactions are always the same and have the form 
 (\ref{phase-shift}). The drawback of free a.q. method is that here we work with numbers directly without even application of numerical notations that generally speaking leads to exponential cost in the computational recourses in comparizon with the bound a.q. method
footnote{The method of free a.q. could be applied if the accuracy of the amplitudes is not important in comparizon with the determining of such basic states for which it is not negligible, in other words when the state in each time instant has the following form
$
\sum\limits_{j\in J}\la_j|\Psi_j\rangle,
$
where the total number of possible states $J$ is limited indepemdently of a time instant.}.
The bounded a.q. thus represent the algebraic form of free a.q., and in what follows we use just the bound for as the most convenient for the notations. 
\nnn

{\Large{Interaction between a particle and a harmonic oscillator}}
\nnn

We consider as an example of a.q. approach the standard problem of a harmonic oscillator interacting with a particle.
This task is important because it represent the model of interaction between charged particle and electromagnetic field. Lagranjian of a system "particle+field" has the form (see (\cite{FH}):
\begin{equation}
L=\frac{mx'^2}{2}-V(x,t)+\frac{MX'^2}{2}+\w^2X^2+g(x',x,t)X(t),
\label{interaction}
\end{equation}
where $x$ and $X$ are the coordinates of a particle and an oscillator, $V$ is the potential energy of a particle. 
We apply the a.q approach to this problem. 
It requires the answer to the following question. How to make agree the coordinates of particle and oscillator when a.q. collide, if $x$ and $X$ are the coordinates of the corresponding a.q., and we cannot require their equality in the collisions ? The simplest solution is as follows. An amplitude quantum for the oscillator has a coordinate of the form  $X_0+X$, where $X_0$ determins its relative spatial position only in the simulating space and the moments of its collisions with the other a.q. and does not participate in the reactions and $X$ is taken from the Lagranjian and it participates in the reactions, where $|\D X_0|\gg |X|$ in each time instant (the swing of pendulum is negligible in comparizon with the shift of a.q.). Correspondingly, the step of modeling of a.q. of the particle $D t\gg \d t$ much exceeds the same step for the oscillator. 
 
We consider the simulation in the framework of reduced Hilbert formalism. 

The transformation of amplitude part of $j$-th amplitude quantum $\a$ of oscillator in the moment $t'$ of its collision with a quantum of the same type is as follows:
\begin{equation}
\la'^{osc}_ j=\la^{osc}_ j\cdot e^{\frac{i}{h}\d S^{osc}_j},\ \ \d S_j^{osc}=\left[\frac{M(X(t')-X(t_0))}{2(t'-t_0)}+\w^2X(t')^2+g\left(\frac{x_j(t')-x_j(t_1)}{t'-t_1},x_j(t'),t'\right)X_j(t')\right]\ dt,
\label{evol-amp-osc}
\end{equation}
where $t_0$ is the moment of previous collision of the quantum $\a$ with a quantum of the same type, $x_j$ 
is the coordinate of the amplitude quantum $\b$ of the particle that is coupled with $\a$, and $t_1$ is the moment of the last collision of $\b$ with a quantum of the same type. 

A transformation of the amplitude part of a.q. of the particle looks similarly. 

We now show how the simulation looks in the "chemical" formalism.  
Here the reactions in the collisions of a.q. of the same type: particle-particle and oscillator-oscillator will be as above.
But to introduce the interaction particle-oscillator we need the supposition about the shift of oscillator itself in the space of cordinates of a particle, e.g. the dynamics of $X_0$. This is the serious question with the physical sense and it arrises in the reduced formalism of Hilbert spaces as well, because the law of movement of a.q. before the coupling is unclear. It shows that the problem reguires the additional conditions that touche the movement of oscillator.
An oscillator cannot be considered as we considered a particle in the potential because it is a carrier of the field itself. This is the conventional approach in the field theory: an oscillator is one mode of an electormagnetic field. We then must assume that a.q. of oscillator is emitted by charged particles (see below). Let we are given the law of movement of oscillator a.q. in the space of coordinates $X_0$. The interaction between a.q. of particle and oscillator takes place in their collisions only. 
If an amplitude quantum $\a$ of the oscillator collides with a quantum $\b$ of the particle we can agree that the reaction goes accordingly to the formula (\ref{evol-amp-osc}), and the reaction for the amplitude part of $\b$ has the same form. Since the summand of interaction $g\left(\frac{x_j(t')-x_j(t_1)}{t'-t_1},x_j(t'),t'\right)X_j(t')$ occurs twice we could put $1/2$ before the coefficient - it can be included to the existing coefficient $g$. It gives the algorithmic reduction of Hilbert formalism if we compose the complex amplitude quanta for many body problem along the method described above.

 If there are several oscillators and they do not interact then the transformations of the ampltude parts of a.q. have the similar form if we take into account the different frequences $\w$; the case of interacting oscillators can be reduced to the case of not interacting by the cange of the coordinate system (see (\cite{FH})), or to write for the interacting oscillators the transformations analogous to (\ref{evol-amp-osc}).

We note that to organize the collisions between the a.q. of the particle and oscillator we need the special assumption about the movements of oscillator a.q., e.g., how $X_0$ varies. Here we asume that a.q. moves chaotically, such that the change of $X_0$ guarantees the number of collisions sufficient for the reaching of the required accuracy. 
\nnn

{\Large{Several charged bodies in the electromagnetic field}}
\nnn

We assumed above that the bodies have the nonzero masses. This consideration can be applied to the case of scalar Coulomb field as well. But if we try to include the separaet photons to the Lagranjian then we would meet the certain difficulties because photons do not disperse the field but they carry it. It requires the radically different approach based on the mein law of electrodynamics - the Maxwell equations. The a.q. approach must be sufficiently flexible that it can be extended to photons. In this section we trace this extension, using the considered problem of interaction between a particle and a harmonic oscillator. We consider a system of charged particles with an electromagnetic field.
The case of many particles is obtained from the case of one particle by the forming of the complex a.q. for many particles and permutations of equivalent particles as was shown above. The specificity of consideration with an electromagnetic field is thus revealed already in the case of one particle + field. This case is represented as a particle interacting with a system of harmonic oscillators which represents a field. This passage needs one particular agreement resulted from the Maxwell equations and which we must assume because this is the agreement that the value of the vector potential of a field is obtained by the summing of the harmonic oscillator coordinates. Photons are quantum of an electromagnetic field, and we must apply our collision model to the photon a.q. that gives the classical explanation of the quantum probabilities. 
But we should consider photon a.q. accounting the photons specificity - as a system of harmonic oscillators interacting with a particles, the more so as the expansion of a field to photons takes place in the impulse representation of the state space but not in the coordinate representation. 
Following our rules from the previous section we asuume that the photon a.q. move so fast that the big number of them have visited the vicinity of a given fixed point in the time frame $\D t$ when an adrone shifts on one step such that we can sum these a.q. and expand the field to photons. 
We consider a system of charged particles with the density $\rho$ in an electric and magnetic fields with field strengths  $E$ and $B$ correspondingly. We define the density vector of a charge $e$ in a point $R,t$ in its shift aAlong the curve $q(t)$ as $j(R,t)=eq'(t)\d^3(R-q(t))$, where $\d^3$ is the three dimension delta-function. 
The main law of evolution for such a system is the system of three Maxwell equations and the equation of a charge conservation:

\begin{equation}
\begin{array}{ll}
\nabla\ E &=4\pi\rho,\\
\nabla\ B &=0'\\
\nabla\times E&=-\frac{1}{c}\frac{\partial B}{\partial t},\\
\nabla\times B&=-\frac{1}{c}\left(\frac{\partial E}{\partial t}+4\pi j\right),\\
\nabla j&=-\frac{\partial\rho}{\partial t}.
\end{array}
\label{max}
\end{equation}

Here the vector and scalar potentials of electromagnetic field can be obtained from the equation
$$
E=-\nabla\phi-\frac{1}{c}\frac{\partial A}{\partial t}.
$$
We consider the impulse representation of magnitudes participating in the Maxwell equations:
\begin{equation}
\begin{array}{ll}
A(R,t)&=\sqrt{4\pi}c\int\bar a_ke^{ikR}\frac{d^3k}{(2\pi)^3},\\
\phi(R,t)&=\int\phi_k(t)e^{ikR}\frac{d^3k}{(2\pi)^3},\\
j(R,t)&=\int j_k(t)e^{ikR}\frac{d^3k}{(2\pi)^3},\\
\rho(R,t)&=\int\rho_k(t)e^{ikR}\frac{d^3k}{(2\pi)^3}.
\label{ext}
\end{array}
\end{equation}
We can agree that (see (\cite{AB, FH}) $\bar a_k=(a_{1,k},a_{2,k})$ is the expansion of vector $a_k$ to two components orthogonal to $k$; the corresponding directions are called the directions of polarization. We assume that these directions are chosen for each vector of impulse $k$ arbitrary and fix this choice. 

The action for such a system is defined as $S=S_{particles}+S_{field}+S_{int}$, where:
\begin{equation}
\begin{array}{ll}
S_{particles}&=\int\sum\limits_j\left(\frac{mq'^2_j}{2}+\sum\limits_l\frac{e_je_l}{|q_j-q_l|}\right)dt,\\
S_{field}&=\frac{1}{2}\int (a'^*_{1,k}a'_{1,k}-k^2c^2a^*_{1,k}a_{1,k}+a'^*_{2,k}a'_{2,k}-k^2c^2a^*_{2,k}a_{2,k})\frac{d^3kdt}{(2\pi )^3},\\
S_{int}&=\sqrt{4\pi}\sum\limits_j\int(a_{1,k}q'_{1,j}+a_{2,k}q'_{2,j})e^{ikq_j(t)}\frac{d^3kdt}{(2\pi )^3},
\end{array}
\label{S}
\end{equation} 
where $q_{1,j},q_{2,j}$ are the projections of the vector $\bar q$ to the dirctions of polarization. The quantum evolution of the considered system can be obtained by the formulas (\ref{wavefunction}),(\ref{Ker}), if we take the sum of actions determined by (\ref{S}) in place of $S$.

A state of our system is represented in the form of a bubble $\cal B$, filled by a.q. of two different types: a.q. of a particle and a.q. of photons of vector field\footnote{If we want to expand a scalar field to photons as well, that corresponds to the derivation of photons from Maxwell equations (\ref{max}) (see, for example, (\cite{AB}), then we must introduce the photons which carry the scalar field, as was explained above.}. 

We apply to such a system many particle approach described above, taking into account that photons carry the field \footnote{If we deal with the bound a.q. we must assume that they are redistributed as was pointed above: a quantum of vector photon dissappearing at the periphery of a bubble is replaced by a quantum emitted by some of charged particles, where its coordinate and impulse is chosen arbitrarily according to the distribution determined by the wave function. In case of free a.q. this mechanism of photon a.q. reproduction is supplemented with their birth in the collisions with a.q. of a particle. The described scheme makes possible to calculate approximately the real physical values characterizing a field. 
For example, the calculation of the vector potential $A$ in the point $R$ in a time instant $t$ can be done by the following formula
$$ 
A(R,t)=\sqrt{4\pi}c\sum\limits_{\tau\in [t,t+\d t],\ \a(\tau )\in C_{R}}\bar X_{\a (\tau )}a^{ik_{\a (\tau )R}},
$$
which is the translation of (\ref{ext}) to the a.q. language.}.

We associate with each vector of impulse $k$ two mutually orthogonal and orthogonal to $k$ vectors of polarization
$p_{k,1}$ and $p_{k,2}$. We describe a.q. of photons of the vector field, that have some peculiarity conecled with the polarization. An amplitude quantum $\a$ of photon has the amplitude $\la_\a$, the coordinate $X_{0,\a}$, vector of impulse $k_\a$, and the oscillator coordinates $a_{1,\a},\ a_{2,\a}$, that are the complex numbers\footnote{Instead of these coordinates that are connected with the chosen direction of polarization we could use the vector of polarization orthogonal to the photon impulse.}. For the modeling of the system evolution in the electromagnetic field we should at first pass to the impulse representation of a.q. that means, accordingly to our method, the multiplication of their amplitude parts to the phase multiplier $e^{-i\bar k\bar x}$. We then can use the standard collision model with only one correction reflecting the feature of interaction between the field and the particles. 
We assume that the trajectories and impulses of the photon a.q. are not changed in the collisions with each other. 
In other words the impulse $k$ does not participate in the process of change of the photon amplitude (that corresponds to the expression (\ref{S}), and also with that the coordinate representation of a photon wave function has not such a sense as for the massive particles (see (\cite{AB}). 

As earlier, we at first consider our problem in the reduced Hilbert formalism. Let a photon amplitude quantum $\a$ be coupled with a particle amplitude quantum $\b$ with impulse $j=k_{\a}$. The transformation of the amplitude part of  $\a$ in a collision with other photon a.q. has the form 
$$
\begin{array}{ll}
\la'_{\a}&=\la_{\a}\cdot e^{\frac{i}{h}\d (S_{1,\a}+S_{2,\a})},\\ 
\d S_{1,\a}&=\frac{1}{2}\left[\left|\frac{a_{1,\a}(t')-a_{1,\a}(t_0)}{(t'-t_0)}\right|^2
+\left|\frac{a_{2,\a}(t')-a_{2,\a}(t_0)}{(t'-t_0)}\right|^2-k_{\a}^2c^2(|a_{1,\a}|^2+|a_{2,\a}|^2)\right]\ (t'-t_0),\\
S_{2,\a}&=\frac{\sqrt{4\pi}}{(2\pi)^3}\sign (e_{\b})\left(a_{1,\a}\frac{x_{1,\b}(t')-x_{1,\b}(t_1)}{t'-t_1}+a_{2,\a}\frac{x_{2,\b}(t')-x_{2,\b}(t_1)}{t'-t_1}\right)e^{i j\cdot x_{\b}}(t'-t_0),
\end{array}
$$
where $x_{1,\b},\ x_{2,\b}$ are the components of vector $x_{\b}$ along $p_{k,1}$ and $p_{k,2}$; where if 
$\a$ is coupled in a list with the others a.q. the corresponding summands must be added to the element of action. 
We now consider the transformation of amplitude parts of a particle quantum $\b$ in its collision with a.q. of the same type. For example, let it be coupled in the list with $\a$ and with a.q. $\g$ of the other particle. The amplitude part then transformes as:
$$
\begin{array}{ll}
\la'_{\b}&=\la_{\b}\cdot e^{\frac{i}{h}\d (S_{3,\a}+S_{2,\a})},\\ 
\d S_{3,\a}&=\frac{m\sign (e_{\b})(x_{\b}(t')-x_{\b}(t_1))^2}{2}+\frac{e_\b e_\g}{|x_\b (t')-x_\g (t')|})\ dt,
\end{array}
$$
If a quantum $\b$ is coupled in the list with the others a.q. of particles, then the corresponding summand must be added to the element of action. 

To pass to the "chemical" formalism we must do the changes in the proposed scheme as in the case of a particle and oscillator. 
\nnn

{\LARGE Appendix 2.\\
About the simulation of Lorentz invariance}
\nnn

The starting point of relativity is Lorentz invariance of the laws of Nature, e.g., the conservation of pseudo-Euclidean metric of the space-time in the passages from one inertial frame to the other. Here by the "laws of Nature" we mean the events that happen on some segment of a pointwise particle trajectory in the space-time with coordinates $x,y,z,t$. If we express such laws by differential equations, we assume that this segment is very small comparatively with the length of a trajectory, and its coordinates are $dx,dy,dz,dt$. Pseudo-Euclidean metric is determined as $ds^2=dx^2+dy^2+dz^2-dt^2$ (we choose the system of units so that the speed of light equals 1. 
Lorentz invariance then means that if we denote by primed variables the values of the corresponding magnitudes in the other inertial frame, then the following equality is true: $ds^2=ds'^2$. In order to consider how this fact can be represented in the algorithmic approach it is required to define the computational network which plays the role of inertial frame. It is done in the next section. 

\subsection{Multihead Turing machines}

We assume the formalization of algorithms in the form of multihead Turing machines\footnote{Markov normal algoritms response to our idea as well. Cellular automata are not appropriate because it do not allow to simulate quantum non-locality.}.

We preceed with the definition of multihead Turing machines. Such a machine consists of three objects: a set of tapes divided into cells, a set of heads and a set of rules for heads shifts which have the form:
$$
a_{j_1},a_{j_2},\ldots ,a_{j_l}\ ;\ q_{k_1},q_{k_2},\ldots ,q_{k_l} \ \ar\ a_{j'_1},a_{j'_2},\ldots ,a_{j'_l}\ ;\ q_{k'_1},q_{k'_2},\ldots ,q_{k'_l}\ ;\ S_{r_1},S_{r_2},\ldots , S_{r_l}
$$
where $a_{j_t},\ q_{k_t}$ denote the contents of cell observed by $t$-th head and the state of this head before the application of the rule, the primed symbols denote these values after the application of the rule, and $S_{r_t}$ denotes the shift which has to be done aver this head, it takes a value from: shift to right, shift to left, no shift. We can launch several Turing machines on the same set of tapes and make the rules for them dependent not only of cells contents and heads conditions, but of what heads of what other machines observe these cells. 
So complicated rules have the following form:
$$
a_{j_1},a_{j_2},\ldots ,a_{j_l}\ ;\ q_{k_1},q_{k_2},\ldots ,q_{k_l} \ ;\ \bar x_{h_1},\bar x_{h_2},\ldots ,\bar x_{h_l}\ \ar\ a_{j'_1},a_{j'_2},\ldots ,a_{j'_l}\ ;\ q_{k'_1},q_{k'_2},\ldots ,q_{k'_l}\ ;\ S_{r_1},S_{r_2},\ldots , S_{r_l}
$$
where $\bar x_{h_t}$ is a list consists of the pairs of the form: (a head number, the corresponding machine number) for all heads observing the cell which is observed by $t$-th head of the considered machine. The different machines will thus interact. We can assume that each many particle amplitude quantum corresponds to exactly one Turing machine which number of heads equals to the number of entangled particles. The number of heads does not influence to our conclusions but we can assume that each machine has two heads only that corresponds to the hierarhical model for many particles systems. All the machines will thus have the same rules that are defined by the interactions between the particles. It is easy to describe the quantum non-locality in terms of multihead Turing machines because it is contained in the rules of machines. For example, for a pair of entangled photons each head points to the spatial location of their amplitude quanta that form a pair. When the local conditions lead to the elimination of one of such quanta, we do not need the special "kill signal" speading from this quantum to its counterpart; the elimination of the both quanta is guaranteed by the application of rule. The realization of rules for multihead Turing machines is the job of the administrative segment. 

Multihead Turing machines give the single treatment of a simultaneity of events in quantum physics that results from the entanglement of the particles. Such a simultaneity consists in the application of a rule to a set of spatially distant heads. Perhaps there are no other simple way to introduce a simultaneity in quantum formalizm. 

\subsection{Why quadratic number of steps is required for simulation}

Every inertial frame can be represented as a set of multihead Turing machines. Its common memory is thus a model of space in this frame. This frame is used as a gage rod for the measuring of the dynamics of objects that lie beyond this frame, for example, particles moving relatively to it. This gage rod physically is the solid object consisting of atoms with fixed positions. Given two such frames which move relatively to each other with the constant speed, the conservation of pseudo-Euclidean metric means the rule of agreement between two frames in the description of the same process in the both frames. Such rules must account the algorithmic description of a dynamics in the both systems, in particular the limitation on the maximal permissible speed of particles in the user segment; the classical law of adding of speeds is thus not applicable here. The rule of agreement must not be classical. We show one argument for that this rule must give the conservation of pseudo-Euclidean metric.  

It was mentioned above that the formal notion of simultaneity for Turing machines is not physically adequate. A simultaneity takes place for the cells observed by heads only in the moment of application of some rule. For the time reckoning in a given inertial frame some standard physical process can be used, for example, the flight of photon through the chain of atoms disposed along the considered trajectory. Since it is described by means of quantum physics, for example, by amplitude quanta, in the modelling of time frame $dt$ we must consider all pairs of points on the trajectory (and even in some vicinity of it which thickness is fixed and independent of its length). 
Each of such pairs corresponds to starting and final points of some amplitude quantum used for the re-count of the wave function in the next time instant\footnote{It does not contradict to that the speed of all photons is the same. The speed of photons arises after the interference of all amplitude quanta only. Here some of them can interact with atoms as was described in the Apendix 1.}. For the simulation of internal processes in our frame in the time $dt$ we must use all states of two head Turing machine acting on the tape of size $dt$, e.g., of the order $dt^2$ elementary operations. 

We now consider the process of the observation the events that happen with particles which are not contained in this gage rod. In this process the pairs of heads will arise which are located a distance of the order of $dS=\sqrt{dx^2+dy^2+dz^2}$ one from another. For example, if such an atom which is disposed outside our gage rod emits a photon, the different photom a.q. are emitted when this atom occupies the different positions relatively to our gage rod. 
Here all external particles for our gage rod are represented by the same cells on the tapes as the internal parts of the gage rod itself. For the simulation of all processes: external and internal it is required the total number of elementary operations of the order $dS^2$. Here the coefficient does not depend on $dt$ and is determined by the starting and final moment of the simulation, but not by the internal time of this frame. If we want to know how many computational steps we have for the simulation of the external pracesses we must subtract the number of operations required for the simulation of internal processes from the total number of operations. We such obtain the value of the form 
$c_1dS^2-c_2dt^2$. We then can imagine that this number of steps required for the simulation of external (e.g., measured) system is the measure of the complexity of its description in this frame. And the equivalence of inertial frames then means that this measure of complexity must be the same for all inertial frames, that gives the conservation of pseudo-Euclidean metric in passages from one frame to the other. The quadratic dependence of the quantity of steps in algorithm from the physical values of the length and the time thus results from the method of the calculation of wave function through Feynman pass integrals if we apply for this calculation the discretization by amplitude quanta. 

This description is not rigorous, let alone to pretend to be the single possible. We have represented it in order to show that the algorithmic approach does not contradict to  Lorentz invariance of the physical laws. 

\end{document}